\documentclass{iopart}[12pt] 
\usepackage{iopams} 
\usepackage[dvips]{graphicx} 
\usepackage{xspace} 
\usepackage{amssymb,amsfonts,amsthm} 

\eqnobysec 

\def\Journal#1#2#3#4#5{(#1) ``#2'' {#3} {\bf #4} #5}  
 
\def\JHEP{\em J. High Energy Phys.} 
\def\CQG{\em Class. Quantum Grav.}

\def\GRG{\em Gen. Rel. Grav.} 
 
\def\CMM{\em Commun. Math. Phys.} 
 
\def\RMP{\em Rev. Mod. Phys.} 
 
\def\JMP{\em J. Math. Phys.}

\newtheorem{theorem}{Theorem}[section] 
\newtheorem{proposition}{Proposition}[section] 
\newtheorem{corollary}{Corollary}[section] 
\newtheorem{lemma}{Lemma}[section] 
\newtheorem{definition}{Definition}[section] 
\newtheorem{example}{Example}[section] 
\newtheorem{remark}{Remark}[section] 
 
\def\Pr{{\it Proof :} \hspace{3mm}} 
\def\qed{\hfill \raisebox{1mm}{\framebox{\rule{0mm}{1mm}}}} 
\def\g{\gamma} 
\def\G{{\bf g}} 
\def\h{{\bf h}}

\def\d{\partial} 
\def\fr{\frac} 
 
\def\k{\vec{\boldsymbol k}} 
 
\def\u{\vec{\boldsymbol u}} 
\def\v{\vec{\boldsymbol{v}}} 
\def\e{\vec{\boldsymbol e}} 
\def\xiv{\vec{\boldsymbol{\xi}}}  
 
\def\bfeta{\boldsymbol{\eta}} 
 
\def\O{\boldsymbol{\Omega}}  
\def\T{\boldsymbol{ T}}

\def\DP{{\cal DP}}

\newcommand{\bd}{\begin{definition}}                
\newcommand{\ed}{\end{definition}}                  
\newcommand{\bc}{\begin{corollary}}                 
\newcommand{\ec}{\end{corollary}}                   
\newcommand{\bl}{\begin{lemma}}                     
\newcommand{\el}{\end{lemma}}                       
\newcommand{\bp}{\begin{proposition}}            
\newcommand{\ep}{\end{proposition}}                
 
\newcommand{\bere}{\begin{remark}}                  
\newcommand{\ere}{\end{remark}}                     
\newcommand{\bex}{\begin{example}}            
\newcommand{\eex}{\end{example}}                
 
\newcommand{\bt}{\begin{theorem}} 
\newcommand{\et}{\end{theorem}} 
 
\newcommand{\be}{\begin{equation}} 
\newcommand{\ee}{\end{equation}} 
 
\newcommand{\bnr}{\begin{eqnarray*}} 
\newcommand{\enr}{\end{eqnarray*}} 
 
\newcommand{\bit}{\begin{itemize}} 
\newcommand{\eit}{\end{itemize}}

\newcommand{\N}{\ensuremath{\mathbb{N}}\xspace}     
\newcommand{\R}{\ensuremath{\mathbb{R}}\xspace}     
\newcommand{\LLL}{\ensuremath{\mathbb{L}}\xspace}      
\newcommand{\SSS}{\ensuremath{\mathbb{S}}\xspace}

\begin{document} 
 
\title[Further properties of causal relationship:...] 
{Further properties of causal relationship: causal structure stability, 
new criteria for isocausality and counterexamples} 
\author{\dag Alfonso Garc\'{\i}a-Parrado and \ddag Miguel S\'anchez} 
 
 
\address{\dag\ Departamento de F\'{\i}sica Te\'orica, Universidad  
del Pa\'{\i}s Vasco, Apartado 644, 48080 Bilbao (Spain)} 
 
\address{\ddag\ Departamento de Geometr\'{\i}a y Topolog\'{\i}a,  
Facultad de Ciencias, Universidad de Granada,  
Avenida Fuentenueva s/n, 18071 Granada (Spain)} 
 
\eads{\mailto{wtbgagoa@lg.ehu.es} and \mailto{sanchezm@ugr.es}}

\begin{abstract} 
\noindent Recently ({\em Class. Quant. Grav.} {\bf 20} 625-664) 
the concept of {\em causal mapping} between spacetimes  
--essentially equivalent in this context to the {\em chronological map} one in abstract chronological spaces--,  
and the related notion of {\em causal structure},  
have been introduced as new tools to study causality in Lorentzian  
geometry.  
In the present paper, these tools are further developed in  
several directions such as: (i) causal mappings  
--and, thus, abstract chronological ones-- do not preserve two levels  
of the standard hierarchy of causality conditions (however, they preserve 
the remaining levels as shown in the above reference),  
(ii) even though global hyperbolicity  
is a stable property (in the set of all time-oriented Lorentzian metrics  
on a fixed manifold), the causal structure of a globally hyperbolic  
spacetime  can be unstable against perturbations;  
in fact, we show that the causal structures of  
Minkowski and Einstein static spacetimes remain stable,  
whereas that of de Sitter becomes unstable, (iii)  general criteria  
allow us to discriminate different causal structures in some general  
spacetimes (e.g. globally hyperbolic, stationary standard); in particular,  
there are infinitely many different globally hyperbolic causal structures  
(and thus, different conformal ones) on $\R^2$, (iv) plane waves with the  
same number of positive eigenvalues in the frequency matrix share the same  
causal structure and, thus, they  
have equal causal extensions and causal boundaries. 
\end{abstract} 
 
\submitto{\CQG} 
 
\section{Introduction} 
 
Lorentzian (time-oriented) manifolds are naturally equipped with a notion of  
{\em causal structure}. Traditionally this has been related to two concepts:  
(A) the {\em classical causality theory}, based on the binary  relations  
``$\leq$'' (causality) and ``$\ll$'' (chronology)  
from which one  
defines the basic sets $I^+(p), I^-(p), J^+(p), J^-(p)$,  
and (B) the {\em conformal structure} generated 
 by the Lorentzian metric. Even though all these approaches are well settled,  
the elements present in (A) and (B) have specific characteristics, and 
it is not clear if they must be  
regarded as the {\em unique} ingredients in a sensible definition of causal  
structure.  As an extreme example, in any totally vicious spacetime all the  
points are related by ``$\ll$''; but there are many different types of totally  
vicious spacetimes, and it is not evident why all of them should be  
deemed as bearing the same causal structure. On the other hand, the conformal structure  
is very restrictive and,  frequently, it seems reasonable to consider  
some non--conformally related spacetimes as causally equivalent  
(otherwise, the concept of causality itself would mean just conformal  
structure in Lorentzian signature, and would be rather redundant).  
For example, most  modifications of a Lorentzian metric around a point  
(say, any non-conformally flat perturbation of Minkowski spacetime 
 in a small neighbourhood) imply a different conformal structure; but,  
one may have a very similar structure of future and past sets for all points.  
 
Two points $p$, $q$ of a Lorentzian manifold are related by  ``$\ll$'' (resp. ``$\leq$'') 
if they can be joined by a future-directed timelike (resp causal) curve. Hence  
classical causality (a global concept) stems from the Lorentzian cone   
(a local concept) but the passage from the latter to the former is not fully grasped  
by the connectivity properties of the above binary relations as we hope to  
make clear in this paper with examples. Nevertheless, recall that,  
in any distinguishing spacetime, the (too restrictive)  
conformal structure is determined by the (too general)  
binary causal relations induced locally by the metric.  
Thus, to find a concept  which retains 
the essentials of binary causal relations but not reducible (in causally 
well-behaved spacetimes) to the conformal structure, becomes a subtle 
question. This concept should lie somewhere in between the local 
information provided by the light cones and the global character of the 
binary causal relations.

In \cite{CAUSAL} a new viewpoint toward this issue was put forward.  
The idea is to define mappings between Lorentzian manifolds which  
transform causal  
vectors into causal vectors or {\em causal mappings},   
and so they preserve the relations ``$\ll$'' and ``$\leq$''.  
Their possible existence between two spacetimes induces a concept of  
{\em causal equivalence} or {\em isocausality} as well as a partial ordering on 
 the set of all the  Lorentzian manifolds.   
 As causal mappings are more flexible than conformal ones, they are a new invaluable tool   
to address in precise terms what is meant by ``the causal structure'' of the spacetime (see sections 4.2 and 4.3 of \cite{REVIEW} for a summary). 
 
The present paper makes a deeper study of such mappings and the associated  
causal relationships, extending and improving \cite{CAUSAL}  
in several directions: (i) to consolidate the foundations of the theory, settling,  
for example,   
unsolved issues on the relation between causal mappings and causal hierarchy,  
(ii) to discuss new related ingredients, as {\em the stability} of the causal  
structure,  
(iii) to obtain criteria  
(such as obstructions to the existence of causal mappings)  
which make the theory more applicable, and (iv)  
to apply them in some relevant families of spacetimes,  
which include globally hyperbolic ones and pp-waves. 
 
\smallskip 
\smallskip 
 
\noindent This paper is organised as follows: 
in section \ref{essentials}, firstly, the essential properties of causal mappings proven in \cite{CAUSAL} are briefly summarized and revisited. Some clarifying properties  
and examples are provided, such as example \ref{ex-no-tim} on the role of time-orientation,  
example \ref{excoj} on totally ordered chains by causal mappings, or proposition \ref{a-p}, 
which deals with the stability of the existence of a causal mapping.  
Also we put forward the notion of {\em causal embedding boundary} in  
subsection \ref{boundary}.  
Remarkably, in the last subsection the stability of the causal structure  
of Minkowski spacetime $\LLL^n$ is proven by finding a  
generic family of  
isocausal perturbations of the metric   
(theorem \ref{b-t}). The aim of this result is twofold: on one hand, it shows the  
(desirable) stability of the causal structure of $\LLL^n$; on the other hand,  
as these perturbed metrics are generically non-conformally flat, this supports  
the claim that the causal structure is induced from causal mappings and, thus, it is a  
more general structure  
than the conformal one. 
 
In section \ref{versus}, first we connect causal mappings with more abstract approaches  
such as Harris ``chronological mappings'' (which preserve ``$\ll$'', see \cite{HARRIS1}),  
and we show that, even though the former are more restrictive than the latter,  
they become equivalent in causally well-behaved spacetimes, theorem \ref{tharris}. Then, we study  
at what extent the standard hierarchy of causality conditions is preserved by causal mappings.  
Despite being known that most of the conditions of this hierarchy are preserved  
(\cite[theorem 5.1]{CAUSAL}, see theorem \ref{preservation} below),  
the preservation of two levels --causally simple and causally continuous-- remained open.  
We give an explicit counterexample answering the question in the negative.  
We emphasize that this counterexample  
also works for related concepts such as chronological mappings, and the 
implications thereof are discussed.  
 
In section \ref{negative}, new obstructions to the existence of causal mappings between  
two spacetimes are provided. These obstructions have a different nature to  
those presented in \cite{CAUSAL} where all are rooted in the causal hierarchy 
and in proposition \ref{curve} below. So, our new criteria show  
the nonexistence of  
causal mappings between spacetimes belonging to the same level of 
the standard hierarchy  (example \ref{ex4.2}),  
allowing us to find many different causal structures,  
even in spaces as simple as rectangles  (example \ref{ejerectan})  
or globally hyperbolic open subsets of $\LLL^2$  
(example \ref{ex4.3}). In particular, results about the existence of infinitely many different  
simply-connected conformal Lorentz surfaces by Weinstein \cite{WEINSTEIN} are extended. 
 
In the last two sections, a first causal classification of two relevant families of spacetimes  
is carried out. Concretely, in section \ref{c-s} 
a general family of smooth products $I\times S, I\subseteq \R$, which includes both,  
standard stationary and globally hyperbolic spacetimes, is studied.  
Time arrival functions $T^\pm$ (introduced in \cite{SANCHEZ2, BARI}) are shown to be related  
to the existence of {\em particle horizons}    
and, then,  
to obstructions to the existence of causal mappings. 
Among the many results of this section, we highlight the following three:  
(1) a  general criterion on the existence of causal mappings applicable to globally hyperbolic  
spacetimes (theorem \ref{split-condition}),  
(2) a classification of  the causal structures of spatially closed   
generalized Robertson Walker (GRW) spacetimes,  
(theorem \ref{tclasifGRW}), and (3) the {\em instability} of the causal structure of  
de Sitter spacetime (theorem \ref{tdSinest}), in clear difference 
with the stability of its global hyperbolicity, or the stability of $\LLL^n$   
or other GRW spacetimes, as Einstein static Universe. 
 
In section \ref{smp} we consider a general family of metrics including  
the important case of $pp$-waves and we give a general criterion for  
the existence of causal mappings, theorem \ref{mp-result}.  
Then, we focus on plane waves, and among other results, we prove that  
{\em locally symmetric plane waves} are isocausal whenever their frequency matrices have the  
same signature, proposition \ref{quadric-equivalent}. 
In the last subsection, we explain how this approach yields information about  
{\em causal boundaries} of plane waves according to the notion introduced in  
subsection \ref{boundary}. In particular, we prove that in the case of the  
frequency matrix being 
negative definite the plane wave admits a  
{\em causal extension} to $\LLL^n$ and a causal embedding boundary consisting  
of two lightlike planes. 
This holds even if the spacetime  
is not conformally flat, a case never tackled before as far as we know. 
 
\section{Essentials of causal relationship} 
\label{essentials} 
 
\subsection{Basic framework}  
\label{s2.1} 
In next paragraphs we recall basic concepts of \cite{CAUSAL} which will be  
profusely used in this work. Italic capital letters 
$V, W,$ ... will denote  
differentiable $C^{\infty}$ manifolds and eventually we will use  
subscripts $V_1$, $V_2$ or a tilde, $\tilde V$. 
Boldface letters will  
be reserved for elements of the tensor bundle associated to a manifold  
(we use this same 
convention to represent sections of this bundle leaving to the context the  
distinction between each case). 
 The special case of vectors and vector fields  
will be distinguished by adding an arrow to the boldface symbol.  
$(V,\G)$ will denote a  time-oriented Lorentzian manifold with  
metric tensor $\G$ (eventually with subscripts, if there is more than one) but we will sometimes  
abuse of the notation and use only the capital symbol to denote the Lorentzian manifold. 
We choose the signature convention $(+,-,\dots,-)$ which means that a vector 
$\u$ is timelike if $\G(\u,\u)>0$, lightlike if $\G(\u,\u)=0$ and spacelike otherwise. 
Timelike and lightlike vectors are called causal vectors and the  
causal vector $\u$ is future-directed if $\G(\u,\v)> 0$ where $\v\neq\u$ 
is the causal vector defining the causal orientation. As usual, we denote  
$I^+(p)=\{x\in V:p\ll x\},\ J^+(p)=\{x\in V: p\leq x\}$, and analogously for their past duals. 
Smooth maps between manifolds are represented by Greek letters and if  
$\Phi:V_1\rightarrow V_2$ is any of such maps then the push-forward and pull-back  
constructed from it are $\Phi_*\T$ and $\Phi^*\T$ respectively.   
\bd 
Let $\Phi:V_1\rightarrow V_2$ be a global diffeomorphism between 
two manifolds.  
We say that the Lorentzian manifold $V_2$ is 
{\em causally related} to $V_1$ by $\Phi$, denoted 
$V_1\prec_{\Phi}V_2$, if for every causal  
future-directed $\u\in T(V_1)$, 
$\Phi_*\u\in T(V_2)$ is causal future directed too. 
The diffeomorphism $\Phi$ is then called a causal mapping.  
$V_2$ is said to be {\em causally related} to $V_1$, 
denoted simply by $V_1\prec V_2$, if there exists a 
causal mapping $\Phi$ such that $V_1\prec_{\Phi}V_2$. 
\label{causal-map} 
\ed 
 
\bere 
{\em The diffeomorphism $\Phi$ is a causal mapping if and only if the  
lightcones of the pull-back metric $\Phi^*\G_2$ include the cones of $\G_1$,  
and the time-orientations are preserved.  
Thus, for practical purposes, one can consider a single  
differentiable manifold $V$ in which two Lorentzian metrics  
$\G_1$, $\G_2$ are defined and wonder when the cones of $\G_2$ are wider  
than the cones of $\G_1$ (i.e., the identity in $V$ is a causal mapping). 
Some of the forthcoming results  
are formulated in this picture.  
In our exposition we will resort to  
one or other picture depending on what we wish to emphasize in  
each context. 
} 
\ere 
A similar definition in which causal past-directed vectors are mapped 
into causal future-directed ones (anticausal mapping) can also be given.   
All the results described below hold likewise for causal and anticausal 
mappings although we only make them explicit for causal mappings.  
Nevertheless, the existence of a causal mapping does not imply the  
existence of a anti-causal one, nor vice versa  
(example \ref{ex-no-tim} below).  
Causal and anticausal mappings are then characterized by the condition   
\be 
\fl\G_2(\Phi_*\u,\Phi_*\u)=\Phi^*\G_2(\u,\u)\geq 0,\ \ \ 
\forall\, \u, \in T(V_1)\ \mbox{causal future-directed}. 
\label{dp-condition2} 
\end{equation} 
This means that $\Phi^*\G_2$  
satisfies the {\em weak energy condition} which is a well-known algebraic  
condition in General Relativity for the stress energy tensor, but also applicable to any symmetric rank-2 covariant tensor. Alternatively, we  
find that $\Phi$ is causal or anticausal iff 
\be 
\fl\G_2(\Phi_*\u_1,\Phi_*\u_2)=\Phi^*\G_2(\u_1,\u_2)\geq 0,\ \ \ 
\forall\, \u_1,\ \u_2 \in T(V_1)\ \mbox{causal future-directed}, 
\label{dp-condition} 
\end{equation} 
which means that $\Phi^*\G_2$ satisfies the {\em dominant property} 
another of the standard energy conditions used in General Relativity. 
For a general rank 2 symmetric tensor, this condition  
is more restrictive than the weak energy condition but,  
as $\Phi^*\G_2$ is a Lorentzian scalar product at each point,  
both conditions become equivalent here  
(see remark \ref{lorentzian} for further details).  
 
Clearly the relation ``$\prec$'' is a preorder in the set of all the 
diffeomorphic Lorentzian manifolds. 
Other basic properties easy to prove are the following \cite{CAUSAL}. 
\bp[Basic properties of causal mappings] 
If $V_1\prec_{\Phi}V_2$, then: 
\begin{enumerate} 
 
\item All timelike future-directed vectors on $V_1$ are mapped to  
timelike future-directed vectors. If the image $\Phi_*\u$ of a  
 causal vector  
$\u$ is future-directed lightlike,  
then $\u$ is a future-directed lightlike vector. 
 
\item\label{3} Every  future-directed timelike (causal) curve is  
mapped by $\Phi$ to a  future-directed timelike (causal) curve 
(this property characterizes  causal mappings; the timelike curves can be regarded  
smooth or only continuous in a natural sense). 
 
\item For every set $S_1\subseteq V_1$, $\Phi (I^{\pm}(S_1))\subseteq  
I^{\pm}(\Phi (S_1 ))$, $\Phi (J^{\pm}(S_1))\subseteq  
J^{\pm}(\Phi (S_1 ))$, and $D^{\pm}(\Phi(S_1))\subseteq  
\Phi(D^{\pm}(S_1))$. 
 
\item\label{point5} If a set $S_2\subset V_2$ is acausal (achronal),  
then $\Phi^{-1}(S_2)$ is acausal (achronal). 
 
\item If $S_2 \subset V_2$ is a Cauchy hypersurface, then  
$\Phi^{-1}(S_2)$ is a Cauchy hypersurface in $V_1$. 
 
\item $\Phi^{-1}(F)$ is a future set for every future set $F\subset V_2$;  
and $\Phi^{-1}(\d F)$ is an achronal boundary for every achronal  
boundary $\d F\subset V_2$. 
\end{enumerate} 
\label{lista} 
\ep 
The inverse of a causal mapping is not necessarily a causal mapping, in fact:  
\bp 
For a diffeomorphism $\Phi:(V_1,\G_1)\rightarrow(V_2,\G_2)$  
the following assertions are equivalent: 
\begin{enumerate} 
\item $\Phi$ (and, thus, $\Phi^{-1}$) is conformal, i.e., $\Phi^*\G_2=\lambda\G_1$, for some function $\lambda>0$. 
\item $\Phi$ and $\Phi^{-1}$ are both causal or both anticausal 
mappings. 
\end{enumerate} 
\label{phi-phiminusone} 
\ep 
Of course, one can find pairs of Lorentzian  
manifolds $V_1$, $V_2$ such that $V_1\prec V_2$ but $V_2\not\prec V_1$ (this last  
statement means that there is no diffeomorphism $\Phi:V_2\rightarrow V_1$ which is  
a causal mapping). Given two diffeomorphic Lorentzian manifolds  
$V_1$, $V_2$ whether $V_1\prec V_2$ or $V_1\not\prec V_2$  
cannot in principle be solved in simple terms.  In some relevant examples,   
the relation $V_1\prec V_2$  
can be proved by constructing explicitly a causal mapping, 
but specific techniques are needed to prove $V_1\not\prec V_2$.  
This was partly tackled in \cite{CAUSAL}, where explicit examples  
in which causal mappings could not be constructed were provided. In all of  
them, there is a global causal property or condition not shared by the spacetimes, which forbids the existence of the causal mapping in at least one direction. These criteria are essentially contained in the following  two results. 
\bt 
If $V_1\prec V_2$ and $V_2$ is globally hyperbolic, causally stable,  
strongly causal, distinguishing, causal,  
chronological, or not totally vicious, then so is $V_1$. 
\label{preservation} 
\et 
The conditions appearing in this last result comprise most of the  
so-called {\em standard hierarchy of causality conditions}. They  
have been extensively studied in the literature; so, one may  
check by independent methods if these  
conditions are fulfilled. Even more, one can check that the whole scale of ``virtuosity'' between strongly and stably causal spacetimes introduced by  
Carter \cite{CARTER},  
is also preserved in the sense that if $V_2$ is {\em virtuous} to  
the $n$th-degree so is $V_1$ (an brief account of Carter's classification  
can be found in \cite{REVIEW}). Thus, according to  
theorem \ref{preservation} if $V_2$  
meets one of these conditions of the hierarchy but  
$V_1$ does not, we deduce that  
$V_1\not\prec V_2$. 
\bp 
Suppose that there is an inextendible  
causal curve $\g\in V_1$ such that $I^{+}(\g)=V_1$ (resp. $I^{-}(\g)=V_1$)  
but no such curve exists in $V_2$.  
 Then $V_1\not\prec V_2$. 
\label{curve} 
\ep  
The assumptions in proposition \ref{curve} imply that  
any inextendible causal curve in $V_2$ has a {\em particle horizon } 
(future particle horizon if $I^+(\g)\neq V_2$, past particle horizon 
otherwise). The presence of particle horizons for any inextendible causal  
curve is known in simple Lorentzian manifolds.  
Perhaps the most famous example in which this property  
holds is de Sitter spacetime where any inextendible causal curve has  
both future and past particle horizons.  
In this case proposition \ref{curve} implies that there is  
no causal mapping from Einstein static universe to de Sitter spacetime,  
although a causal mapping in the opposite way does exist  
(\cite[example 2]{CAUSAL}; this will be widely extended  
in corollaries \ref{cEidS}, \ref{cesu}).  
We give now another straightforward application, in order to show  
the role of the time-orientation. 
   
\begin{figure}[h] 
\begin{center} 
\includegraphics[width=.45\textwidth]{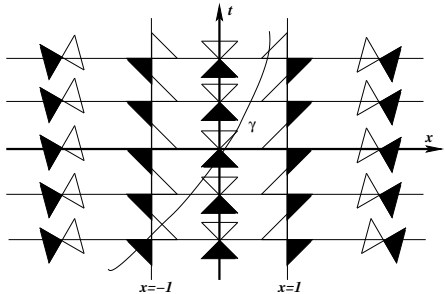} 
\end{center} 
\caption{Example of a two-dimensional spacetime  
with inequivalent causal orientations  
(in this and in the remaining pictures we colour in black the past sheet of  
the causal cone).  
With this causal orientation  any (inextendible) causal   
curve $\g$ has a future particle horizon (this is the line $x=1$ if  
the curve lies in the region $x<1$ or $x=-1$ if the curve lies in the  
region $x>-1$) but no timelike curve has a past particle horizon.  
The spacetime is invariant under translations of the  
$t$ coordinate.}   
\label{inversion} 
\end{figure}

\bex \label{ex-no-tim}{\em  
Consider the spacetime $V$ depicted in figure \ref{inversion}. 
Any inextendible causal curve $\gamma$  
in this spacetime has a future particle horizon.  
Nevertheless, the timelike curve represented by the $t$-axis 
does not have a past particle horizon. 
If the time orientation is reversed, the roles of the future and past horizons  
are interchanged and, thus, there is no causal mapping from the original spacetime  
to the spacetime with the reversed time-orientation, and vice versa. 
This example is analogous to a flat Friedman-Robertson-Walker spacetime with 
no pressure. 
}\eex 
 
\subsection{Causal structures} 
\label{causal-chain} 
We have seen that $V_1\prec V_2$  
can be true despite the spacetimes $V_1$, $V_2$ having rather different  
causal properties (for instance $V_1$ can be globally hyperbolic and  
$V_2$ totally vicious). Things are drastically different if  
$V_1\prec V_2$ and $V_2\prec V_1$, so one defines: 
\bd 
Two Lorentzian manifolds $V_1$ and $V_2$  
are called causally equivalent or 
{\em isocausal} if $V_1\prec V_2$ and $V_2\prec V_1$.  The relation of causal 
equivalence is denoted by $V_1\sim V_2$. 
\label{equivalence} 
\ed    
Traditionally two Lorentzian manifolds have been regarded as ``causally equivalent'' 
when they are conformally related. In principle, this was a sensible point of view because  
conformal transformations put the light cones into a bijective correspondence. 
However, the existence of a conformal relation is a too restrictive assumption 
because, as shown in \cite{CAUSAL}, there are many examples (isolated bodies, exterior of  
black hole regions, etc.) 
in which one may speak of ``essentially equal global causal properties''  
or ``equivalent causality'' but no conformal relation 
exists. These examples are isocausal in the above sense. 
Interestingly enough any pair of Lorentzian manifolds $V_1$, $V_2$ are {\em locally causally 
equivalent}, this meaning that one can choose neighbourhoods of the points $p_1\in V_1$,  
$p_2\in V_2$ which are causally equivalent when regarded as Lorentzian submanifolds  
(see theorem 4.4 of \cite{REVIEW}).     
 
If $V_1\sim V_2$ and one of the  
Lorentzian manifolds complies with the causality conditions stated  
in the theorem \ref{preservation}, then 
so does the other. Therefore the relation ``$\sim$''  
maintains these causality conditions (see subsection \ref{hierarchy} for the  causality  
conditions not included). However, as already explained in  
\cite{CAUSAL} and widely further exemplified below  
there are many non-isocausal spacetimes which lie in the same causality level, i.e.,   
the relation ``$\sim$'' can be used to devise a refinement of  
the standard hierarchy introducing new causality conditions. 
 More precisely,  
the relation ``$\sim$'' is an equivalence relation  
in the set of all the (time-oriented)  
Lorentzian metrics on a differentiable  
manifold $V$, Lor$(V)$. Lorentzian manifolds belonging to  
the same equivalence class  
can be thought of as sharing the ``essential causal structure'',  
and so they deserve their own definition.  
\bd[Causal structure] 
A causal structure on the differentiable manifold $V$ is any  
element of the quotient set {\em Lor$(V)/\sim$}. We denote each causal  
structure by {\em coset$(\G)$} where {\em 
$$ 
\mbox{coset}(\G)=\{\tilde{\G}\in\mbox{Lor}(V):(V,\tilde{\G})\sim(V,\G)\}. 
$$ 
} 
\label{causal-structure} 
\ed 
Now, a partial order $\preceq$ in Lor($V$) can be defined by  
$$ 
\mbox{coset}(\G_1)\preceq\mbox{coset}(\G_2) 
\Leftrightarrow (V,\G_1)\prec(V,\G_2). 
$$ 
This is the natural partial order constructed from the preorder 
``$\prec$''. Causal structures can be naturally grouped in  
sets totally ordered by ``$\preceq$'' in the form  
$$ 
\underbrace{\dots\preceq\mbox{coset}(\G_1)\dots\preceq 
\mbox{coset}(\tilde\G_1)\dots}_{\mbox{glob. hyp.}}\preceq 
\underbrace{\dots\preceq\mbox{coset}(\G_2)\preceq\dots}_{\mbox{causally  
stable}}\preceq 
\underbrace{\dots\mbox{coset}(\G_m)\preceq\dots}_{\dots\,\, \dots}. 
$$  
Of course, some of the groups in a totally ordered chain may be empty;  
for example, if $V$ were compact no chain would contain chronological  
spacetimes. Furthermore the relation ``$\preceq$'' is not a total  
order and so a globally hyperbolic Lorentzian manifold 
need not be related to, say, a causally stable Lorentzian manifold.   
To see this consider the following example  
which also shows that 
even in the case that Lor$(V)$ contain 
globally hyperbolic metrics, such a metric may not exist  
in a totally ordered chain. 
  
\begin{figure}[h] 
\begin{center} 
\includegraphics[width=.9\textwidth]{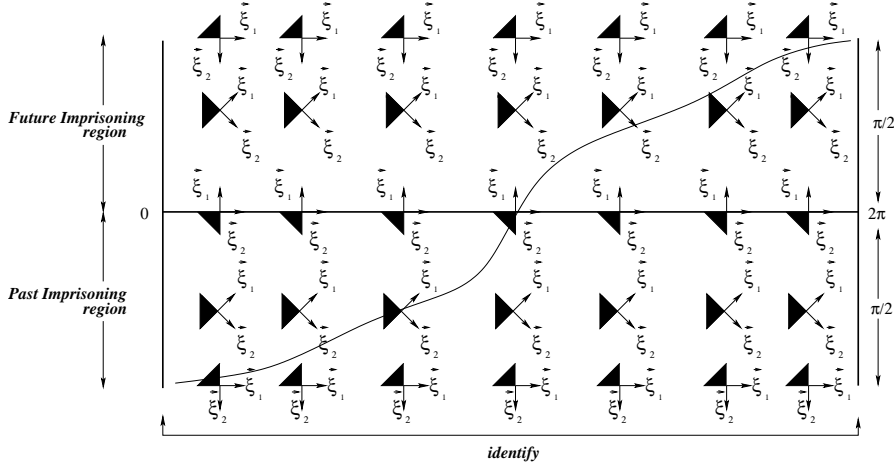} 
\end{center} 
 \caption{This picture represents the  
light cone pattern generated by the metric $\G_2$ of example  
\ref{excoj}. Note the presence of compact regions  
containing totally imprisoned inextendible causal curves (one of 
such curves is shown in the picture).} 
\label{imprisoned} 
\end{figure}

\bex \label{excoj} {\em  
Let the base manifold be a cylinder $V=\R\times S^1$  
and consider the Lorentzian metrics, in natural coordinates, 
$$ 
\G_1= dt^2-d\theta^2,  
$$ 
which is obviously globally hyperbolic, and 
$$ 
\G_2= -\sin (2\varphi(t)) \left(dt^2-d\theta^2\right) +  
2\cos(2\varphi(t))dtd\theta   
\quad \varphi(t) = \frac{\pi}{2} \sin^2t, \forall t\in \R,  
$$ 
which is not chronological, and admits as globally defined  
lightlike vector fields (say, future-directed) 
$$ 
\xiv_1= \cos (\varphi(t)) \partial_t + \sin (\varphi(t))  
\partial_\theta , \quad 
\xiv_2= -\sin (\varphi(t)) \partial_t + \cos (\varphi(t))  
\partial_\theta ,  
$$ 
see figure \ref{imprisoned}.  
Notice that theorem \ref{preservation}  
implies directly $\G_2 \not\prec \G_1$. But in this particular example,   
the converse $\G_1 \not\prec \G_2$ is also true. In fact, note that the 
 structure of the light cones of $\G_2$ imply that any inextendible  
causal curve remains totally imprisoned in some compact subset, say    
$K=[L-\pi/2, L+ \pi/2] \times \SSS^1$ for some  $L\in \R$.   
Thus, if $\G_1 \prec_\Phi \G_2$  
then any timelike curve such as the generatrix $\gamma(t)=(t,\theta_0)$ will  
satisfy that $\Phi\circ \gamma$ is imprisoned in $K$  
and, thus, $\gamma$ must be imprisoned in $\Phi^{-1}(K)$, a  
contradiction. More generally, we can assert: {\em if $\G \in$Lor$(V)$  
contains a non-totally imprisoned causal curve (in particular, if $\G$  
is strongly causal) then $\G \not\prec \G_2$.}} 
\eex   
 
The orderings of causal structures are very appealing  
because they make clear that  
the causal structures defined thereof truly generalize 
most of the standard hierarchy of causality conditions.  
We will delve deeper in this generalization pointing out  
new examples and shedding new light as to the real meaning of  
this generalization.   
 
\subsection{Generalization of the causal boundary} 
\label{boundary} 
 Causal mappings are generalizations of conformal  
relations and thus  
one may expect that most of the concepts involving conformal relations  
can be somehow generalized using causal mappings. One of the most  
fruitful ideas coming up from the conformal techniques  
is Penrose's definition of conformal boundary which allows us to extract  
a wealth of information from spacetimes with good enough causal properties.  
Generalizations of Penrose work have been pursued for many years in  
the literature (see \cite{REVIEW} for a summary of them). 
Following these guidelines a {\em causal boundary} was introduced  
in \cite{CAUSAL}  
by using causal mappings. Here we elaborate on the causal boundary concept of  
\cite{CAUSAL} and put forward the notion of {\em causal embedding boundary}. 
\bd 
Let $i :V\rightarrow \tilde{V}$ be a (non-onto) embedding  
and assume that $i$ is a causal mapping onto is  
image $i(V)$. In this case, when  
$V\sim i(V)$ then $i:V\rightarrow \tilde{V}$  
is a {\em causal extension} of $V$,  
and the boundary $\partial_iV$ is called {\em the causal embedding  
boundary of $V$ with respect to $i$}.  
If, additionally, $i(V)$ has compact closure in  
$\tilde V$, the extension is {\em complete}. 
\label{cboundary} 
\ed 
In the definition of causal boundary presented in \cite{CAUSAL} 
the embedding $i$ was not required to be a causal mapping. 
By imposing this additional condition on $i$,  
we avoid causal extensions not related directly to the original manifold  
$V$. In particular, this would allow us  
to attach  concrete points in the boundary  
to inextendible causal curves $\gamma$ in $V$ with $i\circ \gamma$  
extendible in $\tilde V$, extending properly the classic conformal boundary 
(see definition 6.3 of \cite{CAUSAL}). 
Nevertheless, we must emphasize that (even in the complete case) 
the points in the boundary not necessarily can be reached by curves 
of type $i\circ \gamma$, as explicit examples by Harris  
\cite{HarRef} show (such examples are obtained in the more general ambient 
of chronological spaces, but they are applicable here, as well as 
his discussion on the significance of proper embeddings to 
find boundaries, ib. Subsection 5.3).  
\bex
{\em
Let $V=\{(x,t)\in\LLL^2:-\pi/2<t<0\}$ and take $\tilde V=\LLL^2$ with 
 $dt^2-dx^2$ as the metric for both manifolds. The map 
$i(x,t)=(t\arctan(x)/\pi,t)$ is a causal map from $V$ to $\tilde V$ 
and thus $\tilde V$ is a causal extension of $V$. The set $i(V)$ 
is the interior of a
triangle whose vertices are the points $(-\pi/4,-\pi/2)$, $(0,0)$ and  
$(\pi/4,-\pi/2)$. However, only the point $(0,0)$ and the segment joining
$(-\pi/4,-\pi/2)$ and $(\pi/4,-\pi/2)$ can be reached by causal curves
in $V$.
\footnote{We are indebted to an anonymous referee for this example.}} 
\eex
 
As we see our concept of causal embedding  
boundary depends on the particular causal  
extension. Therefore we should not expect any kind of uniqueness or intrinsic  
property. 
This can be a drawback, but it  already happens in the case of  
the conformal boundary. One of the main differences between  
the causal boundary of \cite{CAUSAL}   
and the conformal boundary is that the former can be very simple to construct  
(see \cite{CAUSAL} for explicit examples) whereas the latter can only be  
computed in few examples. In subsection \ref{ppwaveb} we present  
explicit relevant examples of causal embedding boundaries  
supporting this assertion.  
 
\subsection{Causal tensors and their algebraic characterization} 
\label{causal-tensor} 
Equation (\ref{dp-condition}) tells us that the tensor  
$\Phi^*\G_2$ complies with the dominant energy condition. 
This condition has a natural interpretation in our ambient,  
because it means that the endomorphism canonically associated to  
$\Phi^*\G_2$ preserves the future-directed causal vectors (see below).   
The systematic study of  
this condition and the tensors satisfying it (future  
tensors) have been already performed in a number of references  
\cite{PI,SUP,LIBRODEPETROV,PLEBANSKI} (see also \cite{MCCALLUM,FF}). 
Here we review without proofs the basic facts needed in this work 
referring the reader to previous list of references for more details.    
\bd 
A $m-$covariant tensor $\T\in T^0_m(x)$  
(space of covariant tensors of rank $m$) 
is said to be a future tensor if  
$\T(\u_1,\dots,\u_m)\geq 0$ for 
any set of causal and future-directed vectors $\u_1,\dots,\u_m$.  
Past tensors are defined in the same way replacing ``$\geq$''  
by ``$\leq$''. A tensor is said to be causal if it is either future or past. 
\label{causaltensor} 
\ed 
The set of rank-$m$ future tensors at the point $x$ will be denoted by  
$\DP^+_m|_x$ and the  
whole set of rank-$m$ causal tensors by $\DP_m|_x$ 
(sometimes we will use the notation $\DP^+_m(\G)|_x$ if  
there are more than one metric defined in our manifold or vector space).  
Bundles of causal tensors for all the variants introduced before 
are defined in the obvious way (we drop the subscripts to denote  
these bundles). A very complete exposition of the basic  
properties of causal tensors can be found in \cite{PI}. Among them  
we highlight that $\DP^+_m|_x$ is a pointed convex cone in  
the vector space $T^0_m(x)$.  
An alternative characterization of future tensors is given next  
(see \cite{PI} for a proof). 
\bp 
$\T\in\DP^+_m|_x$ $\Leftrightarrow$ $\T(\k_1,\dots,\k_m)\geq 0$ for any  
set of lightlike future-directed vectors $\{\k_1,\dots,\k_m\}$.  
\label{cri:1} 
\ep 
For our particular case of symmetric rank-2   
tensors,  
notice first that any such $\T$ 
defines a self-adjoint 
endomorphism 
$\hat{\T}$ on $T_x(V)$ by means of 
\be 
\T(\u_1,\u_2)=\G(\u_1,\hat{\T}\u_2). 
\label{2tensor} 
\end{equation} 
If $\T \in\DP^+_2|_x$ then $\hat{\T}$ maps causal future-directed vectors onto causal  
future-directed vectors (future causal-preserving endomorphism) and vice versa.  
 
The algebraic classification of self-adjoint  
endomorphisms is well-known, even though it is more 
involved than when the scalar product is  positive definite 
(see e.g. \cite{FF,ONEILL,HOGLUND}). 
In particular, one obtains for the causal-preserving case:  
\bp 
A self-adjoint endomorphism  
$\hat{\T}$ is causal-preserving  
if and only if either of the following conditions is satisfied 
\begin{enumerate} 
\item $\hat{\T}$ is of Segre type $[1,1\dots 1]$ or its  
degeneracies and the eigenvalue $\lambda_0$ associated to  
the timelike eigenvector is greater than or equal to the absolute value  
of the remaining eigenvalues.  
\item\label{case2}  
$\hat{\T}$ is of Segre type $[21\dots 1]$ or its degeneracies  
and in the decomposition  
$$ 
\hat{\T}=\hat{\T}_0+\lambda\k\otimes{\boldsymbol k}, 
$$  
with $\hat{\T}_0$ a degeneracy of the type $[(1,1)1\dots 1]$ and $\k$ a  
double lightlike eigenvector of $\hat{\T}$ (which is also a lightlike  
eigenvector of $\hat{ \T}_0$)  
the conditions of previous point hold true for $\hat{\T}_0$ plus $\lambda>0$. 
\end{enumerate}  
\label{DP-condition} 
\ep 
 
\bere \label{lorentzian} 
\em 
The symmetric covariant tensor $\T$ constructed from $\hat{\T}$  
fulfills the weak energy condition if and only if $\hat{\T}$ satisfies any of  
the algebraic conditions of proposition \ref{DP-condition} with the difference 
that in point $(i)$ $\lambda_0$ is greater than or equal to  
the remaining eigenvalues (and, subsequently, this modified property is claimed in $(ii)$ for  
$\hat{\T}_0$).   
 
This characterization permits us to check that 
the dominant property and the weak energy condition are  
equivalent for any 2-covariant symmetric tensor $\T$ of Lorentzian signature.  
Clearly, the dominant condition implies the weak condition. 
To prove the converse, assume first that $\T$ satisfies the claimed modification of condition {\em(i)}  
in proposition \ref{DP-condition}. 
In this case, we can find an orthogonal basis of eigenvectors  
for $\hat{\T}$ 
whose eigenvalues are positive since $\T$ has the Lorentzian signature. Therefore the condition  
$\lambda_0-\lambda_i\geq 0$ can be rewritten as $\lambda_0\geq|\lambda_i|$,  
$\forall i=1,\dots,n-1$, as required for a causal tensor, i.e., condition $(i)$ is fulfilled.  
On the other hand, if $\T$ satisfies the claimed modification of $(ii)$, the assumed decomposition of $\T$ implies 
that $\T$ is of Lorentzian signature if and only if so is $\T_0$ and, thus, the problem is reduced to the previous case. 
\ere 
 
\medskip 
 For any symmetric $\T$ we can define: 
$$ 
\mu(\T)=\{\k\ \mbox{null future directed}:\T(\k,\k)=0\}. 
$$ 
In the case that $\T \in\DP^+_2|_x$ then $\mu(\T)$ coincides with the  set of all the future lightlike eigenvectors of $\hat{\T}$, i.e., the so-called {\em set of the canonical null directions}   
for a future causal preserving endomorphism,  
$\mu(\hat{\T})=\{\k $  
lightlike future directed  
$:\hat{\T}\k\propto\k\}.$

Given two metrics $\G, \tilde{\G}$ on $V$, the lightlike cones of  
$\tilde{\G}$ are  {\em strictly wider} than the cones of $\G$  
(i.e., causal vectors for $\G$ are timelike for $\tilde{\G}$)  
if and only if $\G \prec_{id} \tilde{\G}$ and $\mu (\tilde{\G}) = \varnothing$ at each point ($id :V\rightarrow V$ denotes  
the identity map).  
In this case, the relation $\G \prec_{id} \tilde{\G}'$ also holds for all the metrics $\tilde{\G}'$ in some $C^0$ neighbourhood of $\G$; alternatively,  
the identity is stable  
as a causal mapping, in the following sense. 
\bp \label{a-p} 
Fix $\G\in$Lor$(V)$ and let $\tilde{\G}$ be any metric whose light cones  
are strictly wider than the lightcones of $\G$. 
Then for any symmetric rank-2 tensor field $\T$ there exists a function  
$h_0>0$ such that $\G\prec_{id} \tilde{\G} + \T/h$ for any $h\geq h_0$. 
\ep 
That is, the identity remains a causal mapping under small perturbations of $\tilde{\G}$ (and then $\G$) created by any symmetric tensor field $\T$.  
The proof becomes straightforward from the following algebraic lemma. 
\bl 
Let $\T$, $\O$ be two symmetric  
rank-2 tensor of $T_2^0(x)$.  
\begin{enumerate} 
\item  
If  $\O$ satisfies the weak energy condition and  
$\mu(\O)=\varnothing$ then there exists a positive constant $A_0$ such that  
$A\O+\T$ satisfies the weak energy condition, for all $A\geq A_0$. 
\item  
If, additionally, $\O$ has Lorentzian signature then  
$A\O+\T$ belongs to $\in\DP^+_2|_x$ and has Lorentzian signature for large $A$. 
\end{enumerate} 
\label{a-lemma} 
\el 
 
\noindent 
\Pr Notice that, from the condition  $\mu(\O)=\varnothing$, we have $\O(\v,\v)>0$  
for any causal future-directed vector $\v \neq 0$. Choose  a  
fixed future-directed unit timelike vector $\u$, and recall that  
 $\v$ can be written as $\v=\lambda(\u+ \nu \e)$  
where  $\e$ is spacelike and unit,  
($\G(\e,\e)=-\G(\u,\u)$),  
$\lambda\in\R^+$, and $\nu\in [0,1]$.  
Thus taking into account that $\e$ and $\nu$  
vary  
on a compact set, we have, for any $A>0$ and causal future directed $\v$: 
\bnr 
\fl A\O(\v,\v)+\T(\v,\v)=\lambda(\v)^2 
(A \O(\u+\nu \e,\u + \nu \e) + \T(\u+ \nu \e,\u + \nu\e))\geq\\  
\geq \lambda(\v)^2(A L_1(\u) + L_2(\u)) 
\enr 
where the constants $L_1(\u)$ and $L_2(\u)$ gather the   
lower bounds of $\O(\u+\nu \e,\u + \nu \e)$  
and $\T(\u+ \nu \e,\u + \nu\e)$ respectively.  
Furthermore $L_1(\u)$ is a strictly positive quantity due to the condition 
$\O(\v,\v)>0$. Thus, as $\u$ is fixed,  
the tensor $A\O+\T$ satisfies the weak energy condition for any $A>2|L_2(\u)|/L_1(\u)$. 
 
Finally, when $\O$ has the Lorentzian signature, $A\O+\T$ will be also  
Lorentzian for $A$ big enough. So, the last claim becomes straightforward  
from the equivalence between the weak and dominant properties in this case. 
\qed

\subsection 
{Stability of Minkowski spacetime causal structure} 
The simplest Lorentzian manifold in physical and mathematical terms  
is flat Minkowski spacetime $\LLL^n=(\R^n,\bfeta)$, 
$$ 
\bfeta=dt^2-\sum_{i=1}^{n-1}(dx^i)^2 
$$  
in canonical Cartesian coordinates $\{t,x=x^1,\dots,x^{n-1}\}$. 
On physical grounds one would expect that a slight  
``perturbation'' of this metric  should be ``close'' to $\bfeta$  
in its main  properties. This statement needs further clarification about 
what we mean by perturbation and by  
properties close to those of $\bfeta$. Some results in this direction are:  
(i) Geroch \cite[Sect. 6]{Ge} claimed that global hyperbolicity is a stable  
property in the $C^0$ Whitney topology\footnote{This means that, for any globally  
hyperbolic metric ${\G}$ (in particular, $\bfeta$ on $\R^n$) there exist a $C^0$  
neighbourhood of ${\G}$ containing only globally hyperbolic metrics. See for  
example \cite[Ch. 7, sect. 3.2]{BEE}  
for the usual notion of stability and some details on the $C^r$ Whitney topologies.}  
(this property becomes straightforward from the orthogonal splitting proven in \cite{BERNAL2}),  
(ii) Beem and his coworkers proved that geodesic completeness of $\bfeta$ is stable  
in the $C^1$ topology \cite[Proposition 7.38]{BEE}, and (iii) Christodoulou and Klainerman,  
in a landmark result \cite{CRISTODOULOU}, 
proved the non-linear stability of four  
dimensional Minkowski spacetime; i.e., roughly speaking, a perturbation 
 of any initial data set of  
Einstein field equations  
giving rise to four dimensional Minkowski spacetime, will evolve  
into a spacetime similar to Minkowski spacetime in a certain sense 
--and not, say, to a black hole or to a solution with pathological properties.  
 
In this subsection we show by simple means  
a result in the same direction regarding our notion of causal structure.  
In fact, the causality of Minkowski spacetime is preserved by   
quite a long range of perturbations. These perturbations will include neighbourhoods  
of $\bfeta$ in the Whitney $C^r$-topology, and, in this sense, the causal structure is  
{\em stable}. Nevertheless, as we will see in theorem \ref{tdSinest}, this stability  
{\em do not hold, in general, for the causal structure of globally hyperbolic spacetimes}, 
de Sitter being a remarkable counterexample.

Fixing the parallel timelike direction $\partial_t$, consider the  
auxiliary canonical Euclidean product on $\R^n$: 
$ 
\bfeta_R=dt^2+\sum_{i=1}^n(dx^i)^2, 
$  
with associated norm $\parallel \cdot \parallel_R$. 
For any Lorentzian metric $\G$ on $\R^n$ such that $\partial_t$ remains  
(future-directed) timelike, consider the continuous functions  
$\theta_{\mbox{max}}, \theta_{\hbox{min}}: \R^n\rightarrow \R$  
defined as 
\bnr 
\theta_{\mbox{max}}(p)=\\  
\fl=\mbox{max}\left\{ 
\arccos\frac{\bfeta_R(\partial_t,\partial_t+\e)} 
{\parallel\partial_t\parallel_R\parallel\d_t+\e\parallel_R}:  
\e\in T_p\R^n,\ \G(\partial_t,\e)=0,\ \G(\partial_t,\partial_t)=-\G(\e,\e)\right\},  
\enr 
and analogously for $\theta_{\mbox{min}}(p)$, $\forall p \in \R^n$.  
Clearly, $\theta_{\mbox{max}}$,  
$\theta_{\mbox{min}}$ are continuous and take their values in the open interval  
$]0,\pi[$ (for $\G=\bfeta$, $\theta_{\mbox{max}}\equiv  
\theta_{\mbox{min}}\equiv \pi/4$). Now, let $\theta_+ \in ]0,\pi]$  
(resp. $\theta_- \in [0,\pi[$) be  
the supremum (resp. infimum) of the values of $\theta_{\mbox{max}}$  
(resp. $\theta_{\mbox{min}}$). 
 
\bt \label{b-t} 
If $0< \theta_- \leq \theta_+ < \pi/2$, then $(\R^n,\G)$  
 is isocausal to $\LLL^n$. 
 
Thus, the causal structure is stable in the Whitney $C^0$-topology  
(and, thus, in all the $C^r$-topologies). 
\et 
 
\noindent 
\Pr Consider the flat Lorentzian metric $\bfeta_+$ (resp. $\bfeta_-$) on $\R^n$  
such that $\partial_t$ is unit and timelike, and the $\bfeta_R$-angle between $\partial_t$ and any lightlike vector is  
equal to $\theta_+$ (resp. $\theta_-$). Clearly,  $\bfeta$, $\bfeta_+$ and $\bfeta_-$  
are isometric and, by construction, $\bfeta_+ $ (resp.  
$\G$) is obtained from $\G$ (resp. $\bfeta_-$) by opening the lightcones, i.e.,  
$\bfeta \sim \bfeta_- \prec_{id}\G \prec_{id} \bfeta_+ \sim \bfeta$. 
 
For the last assertion, recall that, for any $0<\theta_{-} < \pi/4 < \theta_{+}<\pi/2$,  
the subset of Lor$(\R^n)$ which contains all the Lorentzian metrics with lightcones  
strictly between $\bfeta_-$ and $\bfeta_+$ defines an open neighbourhood of  
$\bfeta$  
in the Whitney $C^0$ topology. 
\qed 
 
Obviously, there are choices of $\G$ satisfying $\bfeta_- \prec_{id}\G \prec_{id} \bfeta_+$ which are not conformally flat and, for such  
choices, no conformal diffeomorphism between $\G$ and $\bfeta$ exist.  
Nevertheless, it is reasonable  
to think that $\G$ and $\bfeta$ behave qualitatively equal from the  
viewpoint of global causality.  
 
Notice that, in particular, the result holds if $\bfeta= \G$ on all  
$\R^n$ but a compact subset $K$ and, thus, if $\G$ is any  ``compact perturbation''  
of $\bfeta$ which preserves $\partial_t$ as timelike. Of course, such perturbed 
metric $\G$ is globally hyperbolic, geodesically complete and asymptotically flat. Therefore  
we are able to state very simply that a perturbation of  
$\LLL^n$ with compact support cannot create regions with  
strange or undesirable causal properties.

\section{Chronological relations and causal hierarchy} 
\label{versus} 
In this section we explore further the interplay between causal mappings and  
two other typical topics of causality theory: mappings between  
{\em chronological spaces} and the standard causal hierarchy.  
 
\subsection{Causal mappings versus chronological relations} 
\label{bad-behaved}    
Any spacetime is a {\em chronological space} in Harris sense \cite{HARRIS1}.  
This is a pair $(X,\ll)$ where $X$ is a set and ``$\ll$'' a binary relation  
with the same abstract properties as the standard chronology relation  
of a spacetime\footnote{Chronological spaces are a generalization of {\em causal spaces}  
in which another binary relation ``$\leq$'' usually called causality relation  
is also present (see \cite{REVIEW} for a review of all these concepts).}. 
Given two such spaces $(X,\ll)$, $(X',\ll')$, a map $\varphi: X \rightarrow X'$ is said  
{\em chronological} iff, for any $x,y\in X$,    
$x\ll y \Rightarrow \varphi(x)\ll'\varphi(y)$.  
>From point ({\em iv}) of proposition \ref{lista},  any causal  
mapping is a chronological  
mapping in Harris sense. Let us see that, if the spacetime has good enough  
causal properties, the converse also holds. 
\bt \label{tharris} 
Let $V$, $V'$ be two spacetimes with $V'$ distinguishing and  
$\varphi: V\rightarrow V'$ a diffeomorphism. Then $\varphi$  
is a causal mapping if and only if it is a chronological mapping. 
\label{causa-distinguishing} 
\et 
 
\noindent 
\Pr The proof relies on the following property \cite[lemma 5.2]{CAUSAL}:  
in a distinguishing spacetime, any curve  
$\g$ totally ordered by the relation ``$\ll$'' is timelike and causally oriented. 
Then, if $\varphi$ is a chronological mapping and $\g$ is a timelike future-directed 
curve in $V$, its image $\varphi(\g)$ is a totally ordered subset of $V'$ by ``$\ll'$'' 
and hence a timelike future directed curve.   
The result is now a consequence of point {\em (\ref{3})} of 
proposition \ref{lista}. The converse is evident. 
\qed 
 
Causal mappings are more restrictive for non-distinguishing  
spacetimes than chronological mappings, and this makes them  more 
useful in certain cases. For example, if $(V',\G')$  
is totally vicious (say,  G\"odel's metric on $\R^4$), and $(V,\G)$ is  
any spacetime with the only restriction that $V$ be diffeomorphic to $V'$ 
(in our example $\LLL^4$ would do)  
then any diffeomorphism from $V$ to $V'$ is a  
chronological relation --but, of course, not necessarily a causal mapping.  
In general, for non-distinguishing spacetimes, the information provided by  
the chronology is small and, thus,  
the concept of causal mapping  may provide more useful information. 
Chronological mappings between causal spaces have been considered many times  
in the literature (see e.g.  
\cite{KRONHEIMER-PENROSE,BUDICSACHS,VYAS,VYAS2,PARK}).

\subsection{Relation with the causal hierarchy} 
\label{hierarchy} 
As we explained in section \ref{essentials} 
 most of the levels of the hierarchy of causality conditions  
are preserved by isocausality.  
Nevertheless, we are going to give a counterexample 
which shows that the two remaining levels, causal continuity and causal simplicity  
are not preserved. We emphasize that, as causal mappings are more restrictive than  
chronological ones, this counterexample also works for chronological mappings 
when Alexandrov's topology for chronological spaces is considered. 
\bex 
\label{causal-example} 
{\em  
Consider in Minkowski 2-spacetime $\LLL^2=(\R^2,\bfeta)$, null 
coordinates $(u,v)$,  
$\bfeta=-2dudv$ with $-\partial_u, \partial_v$ future-directed,  
and let $V$ be the open subset of  
$\R^2$ obtained   by removing $N=\{(u,v):  v\geq -u \geq 0 \}$.  
Clearly, $(V,\bfeta)$ is not causally continuous  
(and, thus, neither causally simple); in fact:  
$$ 
R=\{(u,v) \in V: u<0, v>0\} \subset \bigcap_{k\in \N}I^+(1, -1/k) 
 \quad \hbox{but} \quad  R\cap I^+(1,0)= \varnothing   
$$ 
(see Fig. \ref{ejemplo}). Next, our aim is to construct a second metric $\G$ on $V$,  
with the light cone at each $p\in V$ strictly wider than the cone for  
$\bfeta$ (i.e., $\bfeta \prec_{id}\G$)  
and such that $(V,\G)$ is causally simple. Let 
$$ 
\G= -2dudv + 2f(u,v) du^2 
$$ 
where $f>0$ is defined below. The two globally defined vector fields  
$\xiv_1=-\partial_u- f \partial_v$, $\xiv_2=\partial_v$ are  
lightlike future-directed 
with respect to $\G$ and thus they define the causal cone of $\G$ at  
each point of $V$. Hence  $\bfeta \prec_{id}\G$ because if $f>0$ the causal cone of  
$\G$ contains the causal cone of $\bfeta$. Alternatively we can check that  
$\G\in\DP^+_2(\bfeta)$ by means of the conditions of proposition \ref{DP-condition}. 
To this end we calculate the endomorphism $\hat{\G}$ whose matrix form in the basis  
$\{\d/\d u,\d/\d v\}$ is 
$$ 
\left(\begin{array}{cc} 2f(u,v) & -1\\1 & 0 \end{array}\right). 
$$    
The algebraic type of this  
matrix falls into second point of proposition \ref{DP-condition} 
with $\lambda=f(u,v)>0$.    
 
Now, let $\varphi:[0,1]\rightarrow \R$ be any smooth non-increasing function with  
$\varphi(0) = 1, \varphi(1) = 0$,  
and define: 
$$ 
f(u,v)= \left\{ 
\begin{array}{lll} 
1 & & \hbox{if} \; u\leq 0;\ \hbox{or} \; v\leq 0;\ \hbox{or};\ v\leq u-1 \\ 
\frac{1+v}{u} & & \hbox{if} \; v\geq u> 0 \\  
\left(\frac{1+v}{u}-1\right) \varphi (s_{uv}^2) + 1 & & \hbox{if} \; u\geq v\geq  
\hbox{Max}\{u-1, 0\}  
\end{array}  
\right. 
$$ 
where $s_{uv}= l_{uv}/L_{uv}$ with the following definitions: given the straight line  
$r_{uv}$ which joins $(u,v)$ and $(0,-1)$, then $Q_{uv}= \frac{u}{1-u+v}(1,1)$ is the  
intersection between $r_{uv}$ and the line $u=v$, $S_{uv}= (\frac{u}{1+v},0)$ is the  
intersection between $r_{uv}$ and the $u-$axis, and $l_{uv}$ (resp. $L_{uv}$) is the 
 usual Euclidean distance between $(u,v)$ and $Q_{uv}$ (resp. $Q_{uv}$ and $S_{uv}$) 
 (see figure \ref{ejemplo}). To check that $(V,\G)$ is causally simple, notice that the causal  
future of any point $P=(u_0,v_0)$ (and analogously the causal past) is the region  
of $V$ lying between the integral curves $\gamma_{\xiv_1}$, $\gamma_{\xiv_2}$ of $\xiv_1$, $\xiv_2$ 
through $P$. These integral curves together with the causal future and past  
for different points are depicted in figure \ref{ejemplo4}  being   
clearly seen that the causal future and past of any point are closed sets. 
This implies that $(V,\G)$ is causally continuous too. 
 
\begin{figure}[h] 
\begin{center} 
\includegraphics[width=.48\textwidth]{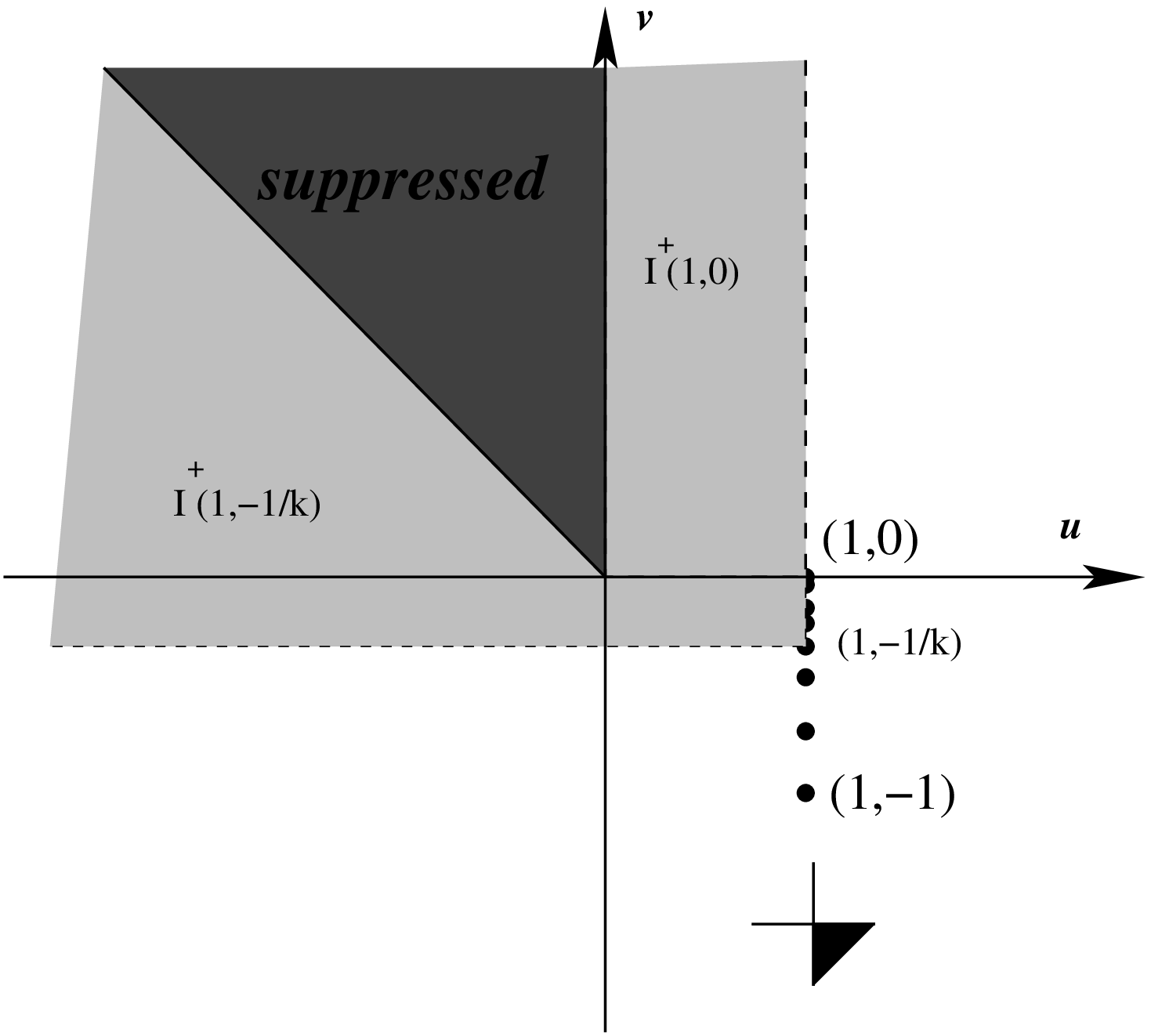} 
\includegraphics[width=.48\textwidth]{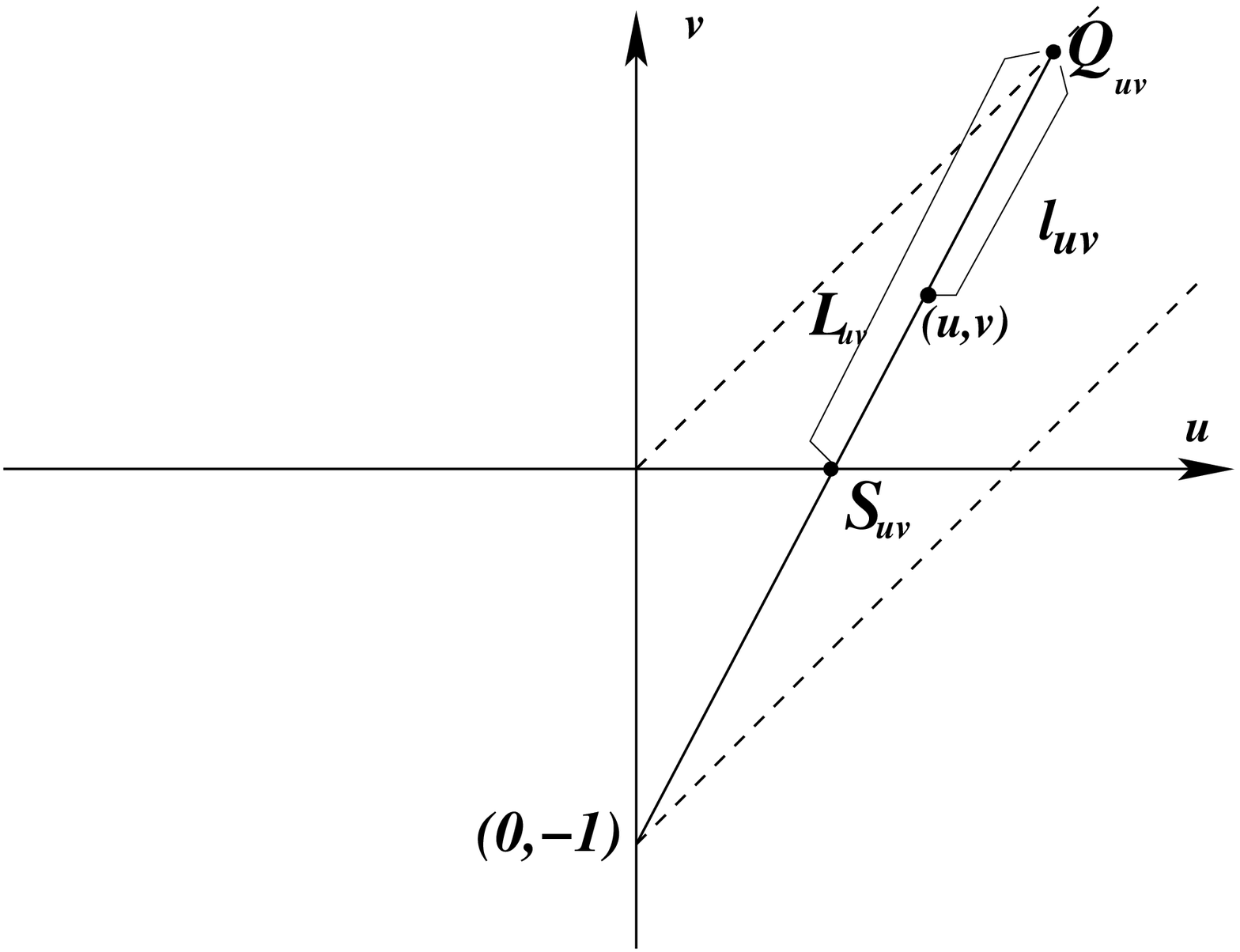} 
\end{center} 
\caption{\footnotesize The left picture is the 2-dimensional spacetime  
$(V,\bfeta)$. 
We have coloured in grey the chronological future of a point of the  
sequence $\{(1,-1/k)\}_{k=1}^{\infty}$. The chronological future of  
$(1,0)$ is the grey region above the $u$ axis {\em and positive $u$}.  
The picture of the right describes the geometric construction 
needed to define $s_{uv}=l_{uv}/L_{uv}$.} 
\label{ejemplo} 
\end{figure} 
\begin{figure}[h] 
\begin{center} 
\includegraphics[width=.48\textwidth]{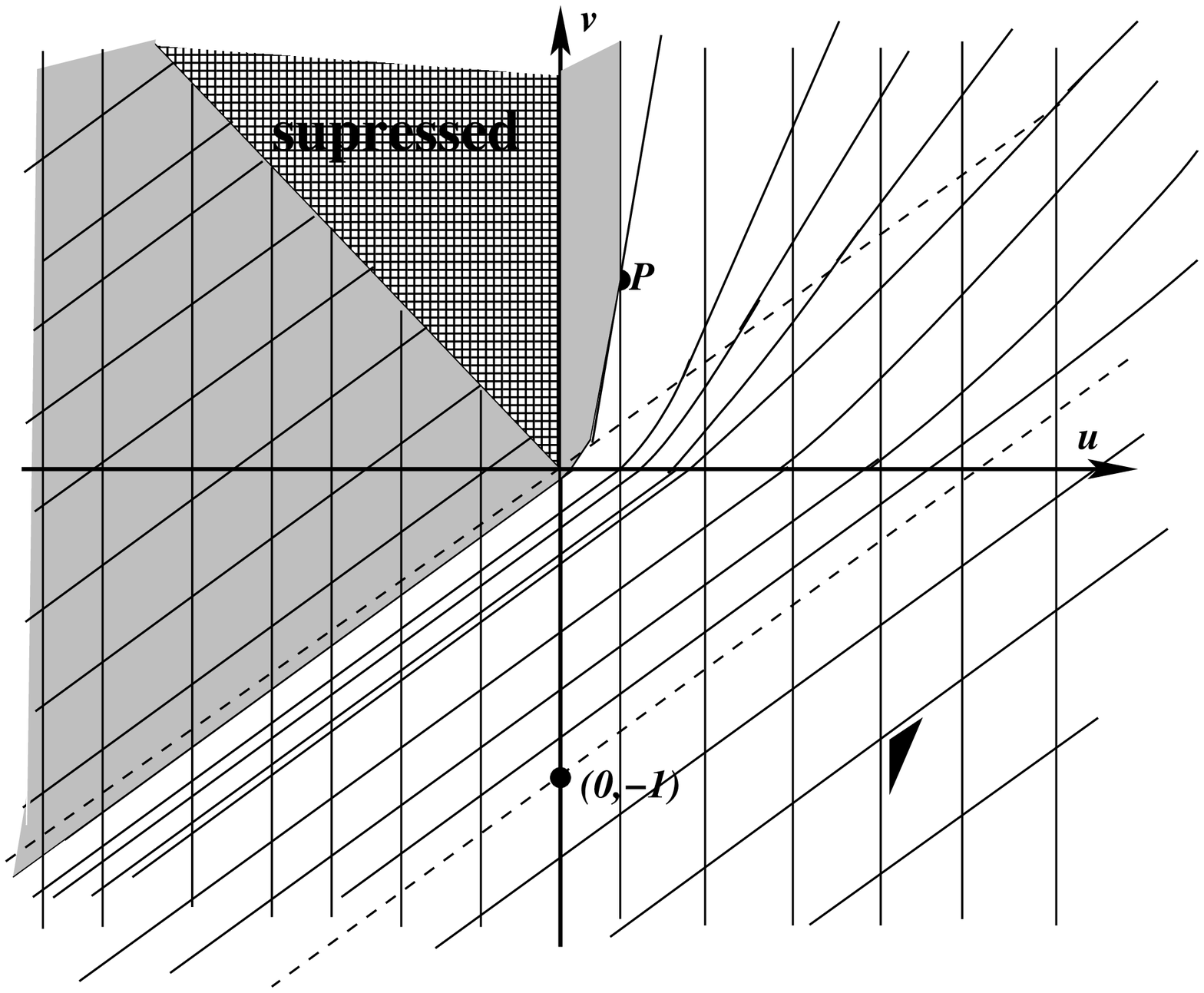}\ \ \   
\includegraphics[width=.48\textwidth]{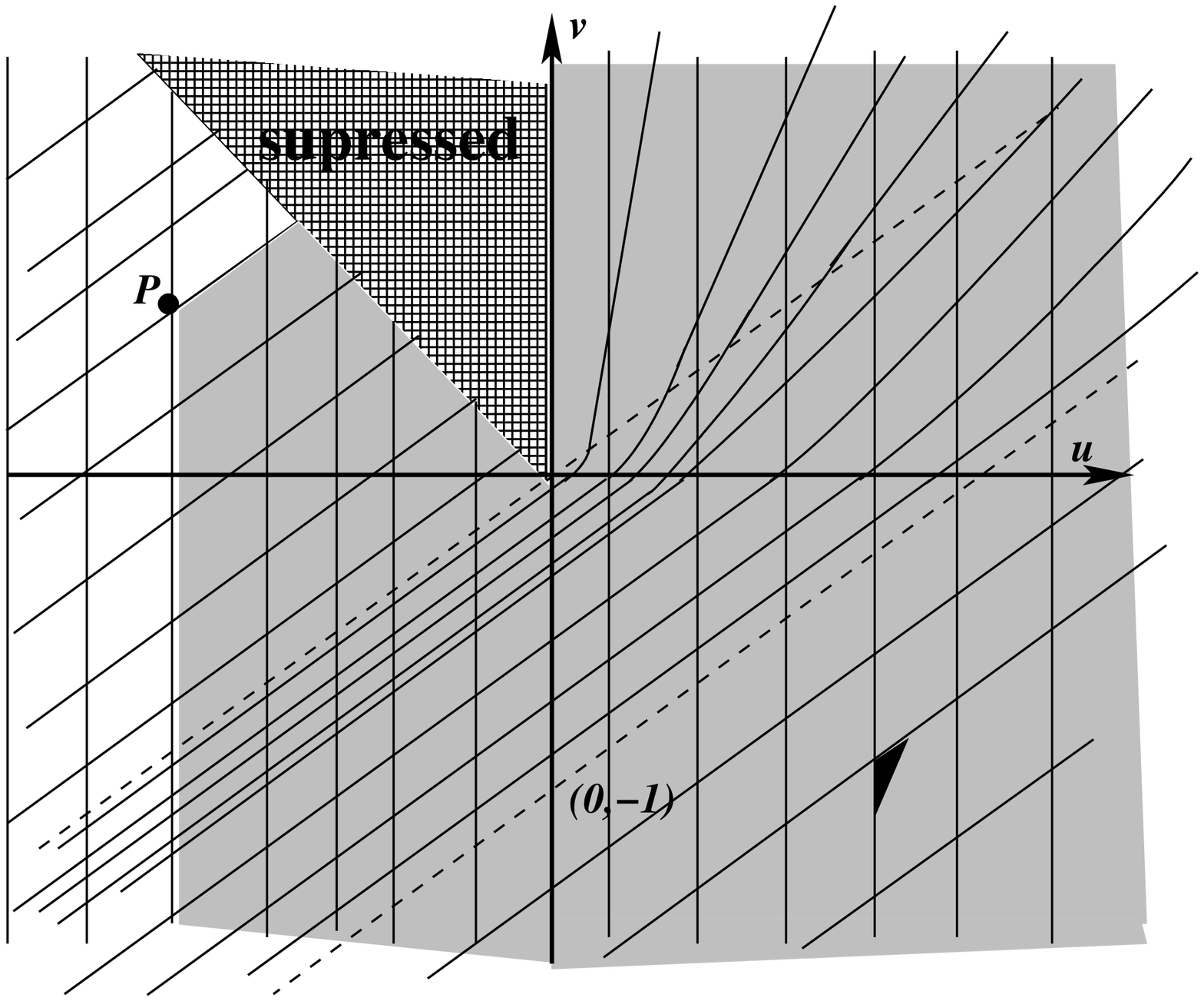} 
\end{center} 
\caption{\footnotesize  In these pictures  
we show the lightlike geodesics of  
$(V,\G)$. The vertical lines  are integral curves of $\xiv_2$ whereas the  
oblique lines are the integral curves of $\xiv_1$.  
In the left picture we have coloured in grey the causal future of a 
point $P$ lying in the first quadrant, whereas in the right picture the grey region corresponds to the causal past of a point $P$ now in the second one. 
In both cases these regions are closed sets. From these pictures  
we deduce that the causal future of any sequence of points approaching 
to $(0,0)$ converges toward the region $u<0$, $v>u$ whereas their causal past  
tends to the region $u>0$.} 
\label{ejemplo4} 
\end{figure} 
}  
\eex 
This example proves that theorem \ref{preservation} does not hold if $V_2$ is causally simple  
or causally continuous. This raises the question as to why these  
two conditions behave differently  
under the action of causal mappings.  
The ultimate reason of this relies on the fact that causality  
conditions covered by theorem \ref{preservation} only deal  with global causal  
properties of the spacetime making them  
causality conditions in a strict sense (to see this observe that they can always  
be formulated in terms of a condition or conditions 
involving only causal curves, see e. g. \cite{FF,BEE,SINGULARITY}). As  
causal continuity and causal simplicity relate causal and topological properties 
of the differentiable manifold,  they are not in the same footing as the other  
conditions\footnote{In spite of the fact that the topology is Alexandrov's one  
(as in any strongly causal spacetime), which is determined purely by the causal relations.}. 
The moral is that although causal continuity and causal simplicity are not covered by theorem  
\ref{preservation},  definition \ref{causal-structure}  should not  
be affected by the existence of two spacetimes with the same {\em causal structure} but  
only one of them being causally simple.

\section{New criteria for non-existence of causal mappings} 
\label{negative} 
In section \ref{essentials} we saw some ways to disprove the existence of causal mappings. They  
involve a global causal property not shared by the Lorentzian manifolds under study and,  
essentially, they were reduced to two criteria:  
the standard causal hierarchy of spacetimes, theorem \ref{preservation}   
(with the limitations pointed in subsection \ref{hierarchy})  
and the nonexistence of horizons,  
proposition \ref{curve}. Additionally, example \ref{excoj} explains a property which can be  
used as a third criterion. 
 
As we are going to see next,  
more elaborate criteria can be used in   
complex situations. The procedure is similar to what we did in section \ref{essentials}: 
we give a number of global causal properties on $V_2$ which are transferred  
to $V_1$ (or vice versa) if $V_1\prec V_2$, and this implies that 
$V_1\not\prec V_2$ if any of these properties fails in $V_1$.

In what follows, any hypersurface $S$ (or submanifold) will be considered smooth, embedded,  
connected and edgeless (thus without boundary).  
Recall that, for a subset of a spacetime $A\subseteq V$, the common past is defined by $\downarrow A\equiv\cap_{x\in A}I^-(x)$.

\bp \label{criterio1} 
Assume that $V_1\prec V_2$ and that $V_1$  
admits $j$ inextendible future-directed causal curves (or, in general,  
$j$ submanifolds at no point spacelike and closed as subsets of $V_1$) $\gamma_i, i=1, \dots , j$ satisfying either of the following conditions: 
 
\begin{enumerate} 
\item $V_1= I^+(\gamma_i) \cup \gamma_i \cup I^-(\gamma_i)$. 
\item $\gamma_i\subseteq\downarrow\g_{i+1}$, $\forall i=1, \dots , j-1$, $j>1$. 
\end{enumerate} 
 
\smallskip 
 
\noindent Then so does $V_2$. 
\label{future-common} 
\ep 
\noindent 
\Pr Denote by $\Phi: V_1 \rightarrow V_2$ the  
causal mapping. From point ({\em iv}) of proposition \ref{lista}   
it is clear that the sets $\Phi(\gamma_i)$, $i=1,\dots,j$ satisfy condition 1   
in $V_2$ whenever $\g_i$, $i=1,\dots,j$ do in $V_1$. To prove the second point  
we have to use the property 
$$ 
\Phi(\downarrow A)\subseteq\downarrow\Phi(A),\ A\subset V_1, 
$$   
which again is a straightforward consequence of  
point ({\em iv}) of proposition \ref{lista}. 
\qed 
 
\begin{figure}[h] 
\begin{center} 
\includegraphics[width=.2\textwidth]{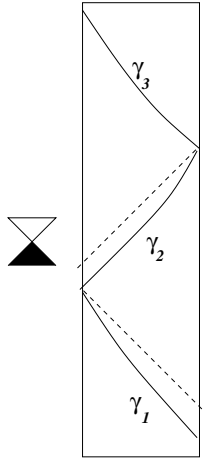} 
\end{center} 
\caption{\footnotesize The curves $\g_1$, $\g_2$, $\g_3$ of this picture meet both  
conditions of proposition \ref{future-common}. The dashed lines are the future boundaries of  
the common past of $\g_2$ and $\g_3$.} 
\label{zig-zag} 
\end{figure}

\bex  \label{ejerectan} 
{\em 
As a simple application, we can show that there are infinitely  
many  rectangles of $\LLL^2$, in  
standard Cartesian coordinates $(t,x)$, which are not isocausal  
(the example is obviously generalizable to $\LLL^n$, by using   
hypersurfaces at no point spacelike instead of causal curves).  
For each $L>0$, let $R_L=\{(t,x): t\in ]0,L[, x\in ]0,1[\} \subset \LLL^2$.  
Note that, if $L< L'$, then $R_L \prec_\Phi R_{L'}$, where  
$\Phi(t,x)= (L't/L, x)$.  
A set of $j$ curves satisfies both properties of proposition  
\ref{criterio1} if and only if $L\geq j$ 
(see figure \ref{zig-zag}).  
Thus, if $L'-L \geq 1$ then $R_{L'} \not\prec R_{L}$. 
This result can be refined if $L=1$ in which case we can actually show that  
$R_{1+\epsilon}\not\prec R_1\not\prec R_{1-\epsilon}$, for any  
$0<\epsilon<1$. To see this note that if $V_1\prec_{\Phi} V_2$ and  
$p\in V_1$ is any point such that $V_1=I^+(I^-(p))=I^-(I^+(p))$ then  
$\Phi(p)$ also satisfies these same properties. In the case of  
$R_1$ only the point $p=(1/2,1/2)$ satisfies previous conditions whereas  
there are infinitely many points for $R_{1+\epsilon}$ (a neighbourhood  
of its centre) and none for $R_{1-\epsilon}$.  
 
} 
\eex 

For the following result, recall that if $\Phi$ is a causal mapping then $\Phi^{-1}$ maps non-timelike vectors to non-timelike vectors. 
\bp 
Assume that $V_1\prec V_2$ and suppose 
that $V_2$ satisfies one of the following properties 
\begin{enumerate} 
\item\label{point1} There exists an acausal (resp. achronal; achronal and spacelike; a foliation by 
 any of previous ones)  hypersurface $S'$, which is closed (resp. compact) as a subset of $V_2$.  
 
\item\label{point2} There exists a hypersurface $S'\subset V_2$ as in (\ref{point1}) such that  
$V_2 \neq I^+(S') \cup S' \cup I^-(S')$.  
 
\item\label{point3} There are $k>1$ hypersurfaces   
 $S'_j\subset V_2$, $j=1,\dots k$ as in (\ref{point1}) 
such that no pair of them can be joined by a causal curve. 
 
\smallskip 
 
\hspace{-.9cm} Then the same property is satisfied by $V_1$. Moreover,  if: 
 
\smallskip 
 
\item\label{point4} all the hypersurfaces in $V_1$ with any of the properties stated in  
(\ref{point1}) are homeomorphic,  
 
\smallskip 
 
\hspace{-.9cm} then so happens in $V_2$.  
 
\end{enumerate} 
\label{set-properties} 
\ep 
\noindent 
\Pr  We prove each case separately (in all cases we take as $\Phi:V_1\rightarrow V_2$  
the diffeomorphism establishing the causal mapping). 
 
\noindent 
{\em(\ref{point1})}\ \  If $S'$ is the stated hypersurface of  
$V_2$ then by point $(\ref{point5})$ of proposition \ref{lista} $\Phi^{-1}(S')$ has these same properties  
as a subset of $V_1$. Moreover if $S'$ belongs to a foliation of $V_2$ then  
$\Phi^{-1}$ gives rise to a foliation of $V_1$ with the required properties. 
 
\medskip 
\noindent 
{\em(\ref{point2})}.\ \  If $V_2\neq I^+(S')\cup S'\cup I^-(S')$ then  
$$ 
\Phi^{-1}(I^+(S')\cup S'\cup I^-(S'))=\Phi^{-1}(I^+(S'))\cup\Phi^{-1}(S')\cup\Phi^{-1}(I^-(S'))\neq V_1. 
$$ 
The result is now a consequence of the property 
$$ 
I^+(S')\supseteq\Phi(I^+(\Phi^{-1}(S'))), 
$$ 
(and analogously for $I^-$) which tells us that $\Phi^{-1}(S')$  
is the sought hypersurface. 
 
\medskip 
\noindent 
{\em (\ref{point3})}.\ \  If no pair of the set $\{S'_{j}\}$, $j=1,\dots k$  
can be joined by a causal curve  
then the same is true of the hypersurfaces $\Phi^{-1}(S'_j)$. 
 
\medskip 
\noindent 
{\em(\ref{point4})}.\ \   
Pick any pair of acausal (resp. achronal, achronal and spacelike) closed (resp. compact)  
hypersurfaces $S'_1$, $S'_2\subset V_2$. The hypersurfaces $\Phi^{-1}(S'_1)$,  
$\Phi^{-1}(S'_2)$ 
are homeomorphic by assumption, and then so are $S'_1$, $S'_2$, since $\Phi$ is a homeomorphism. \qed 
 
Now, we present  some examples showing how to apply conditions of  
this last proposition, and postpone further examples to the next section. 
 
\begin{figure}[h] 
\begin{center} 
\includegraphics[width=.8\textwidth]{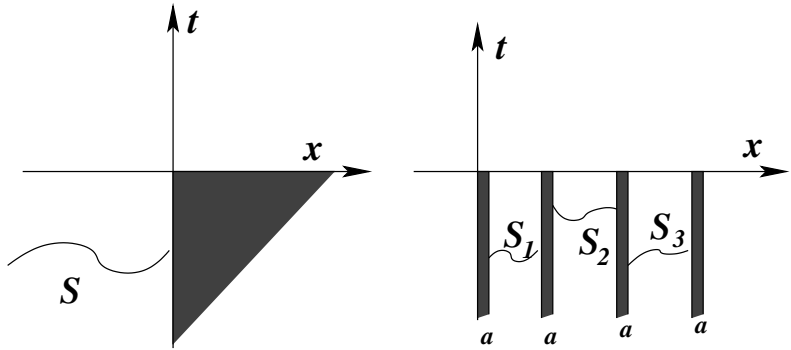} 
\end{center} 
\caption{\footnotesize Examples of spacetimes satisfying properties  
{\em(\ref{point2})} and {\em(\ref{point3})}. 
All the regions in dark grey are suppressed. 
In the left picture $V\neq I^+(S)\cup S\cup I^-(S)$ while in the right picture no pair of the set $\{S_1,S_2,S_3\}$ can be joined by a causal  
curve.} 
\label{prpcd} 
\end{figure}

\bex \label{ex4.2} 
{\em 
A simple spacetime complying with property {\em(\ref{point2})} is  
$\LLL^2$ with any of the quadrants defined by Cartesian coordinate  
axes removed.   
If, instead of a quadrant, we remove the $m$ regions defined by  
$[j, j+a]\times ]-\infty,0]$, $j=0,\dots m-1$, $a<1$ then property  
{\em(\ref{point3})} is satisfied,  
see figure \ref{prpcd}. 
Note that any of these spacetimes (denoted generically by $V$)  
is diffeomorphic to 
$\LLL^2$ but, as $\LLL^2$ does not fulfill either property,  
$\LLL^2 \not\prec V$  
(the opposite causal mapping is also forbidden because  
$V$ is never globally hyperbolic). 
} 
\eex 
 
\begin{figure}[h] 
\begin{center} 
\includegraphics[width=.8\textwidth]{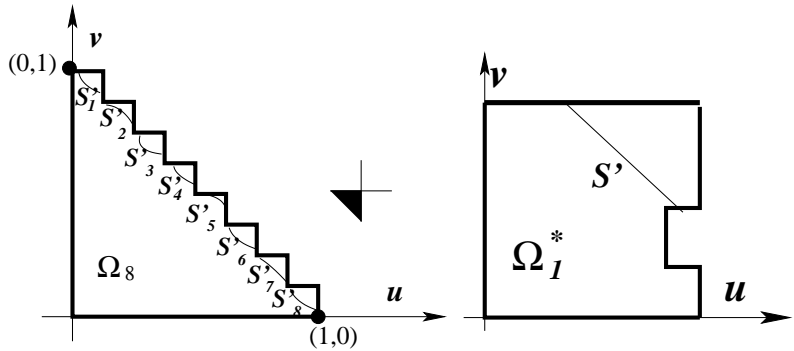} 
\end{center} 
\caption{\footnotesize The picture of the left is the globally hyperbolic set 
$\Omega_8$ where a set of hypersurfaces  
complying with point {\em(\ref{point3})} of proposition \ref{set-properties} has  
been drawn. The picture of the right is $\Omega^*_1$. 
} 
\label{stairway} 
\end{figure}

\bex \label{ex4.3} 
{\em  
Another explicit example of property {\em(\ref{point3})} yields infinitely many globally hyperbolic  
open subsets of $\LLL^2$ which are not isocausal, extending previous results on Lorentz surfaces  
\cite{WEINSTEIN}\footnote{According to Weinstein, a Lorentz surface is a pair $(S,[h])$ where  
$S$ is a (oriented) surface and $[h]$ a (pointwise) conformal equivalence class of Lorentzian metrics  
on $S$ \cite[Sect. 1.3]{WEINSTEIN}.}.  
We again resort to  
$\LLL^2$ but now in null coordinates $\{u,v\}$, $\bfeta=2dudv, 
$ and we define the ``stairway-shaped'' open regions  
$\Omega_{m}$, $m\in\N$ (see figure \ref{stairway}).  
A Cauchy hypersurface can be obtained 
 by drawing a spacelike curve from $(0,1)$ to $(1,0)$.  
All these regions comply with property {\em(\ref{point3})}  
(the set of hypersurfaces $S'_j$ are shown in the picture). The greatest  
number of such hypersurfaces is given by $m$ for each $\Omega_m$ 
so proposition \ref{set-properties} tells us that  
$\Omega_m\not\prec\Omega_{m'}$ if $m<m'$.  
In particular there is no conformal relation  
between $\Omega_m$ and $\Omega_{m'}$ if $m\neq m'$, and  
thus there are infinitely many simply connected Lorentz surfaces  
in the sense of \cite{WEINSTEIN}  
(see this reference for a different proof of this last result).  
Summing up, simple bi-dimensional diffeomorphic globally hyperbolic  
spacetimes with different causal structures are found.  
 
In this same context, consider the manifold $\Omega_1$  
(a square in the plane $u-v$) and define the manifold $\Omega^*_1$  
as the open region of $\LLL^2$ shown in figure \ref{stairway}. 
Clearly there are acausal  
closed hypersurfaces in $\Omega^*_1$ fulfilling property  
{\em (\ref{point2})} (the hypersurface $S'$ of the figure is an example) 
but none in $\Omega_1$ so $\Omega_1\not\prec\Omega^*_1$.  
}  
\eex 
\bex \label{rdS} {\em 
Property {\em(\ref{point4})}  
 is satisfied by de Sitter spacetime $\SSS_1^n$ if  
the hypersurfaces are considered compact (see proposition \ref{pdifSGRW} below),  
but it is not if they are only closed as a subset. In fact, not only does  
$\SSS_1^n$ admit compact spacelike (achronal) hypersurfaces, but also  
 non-compact ones which are closed as a subset of $\SSS_1^n$. 
This can be easily seen if we resort to the  
representation of $\SSS_1^n$ as a unit sphere of 
$\LLL^{n+1}$   with respect to the  
pseudo-distance induced by the Lorentzian metric. The  
intersection of $\SSS_1^n$ with a null hyperplane of $\LLL^{n+1}$  
through the origin is then one of such hypersurfaces. 
 
As we will see in  
the next section, property {\em(\ref{point4})}, even in the case of  hypersurfaces only closed as a subset (but non necessarily compact a priori),  
holds in spacetimes which include  
the standard stationary spacetimes (proposition \ref{pdifS}, remark  
\ref{rdifS}) and some generalizations of Robertson-Walker models  
(proposition \ref{pdifSGRW}). Thus, proposition \ref{set-properties} will forbid the isocausality of any of these spacetimes and $\SSS_1^n$. 
}\eex 
It is not difficult to give other examples which show the   
applicability of previous criteria, as well as to find new criteria by making simple variations  
of proposition \ref{set-properties} (see, for example, remark \ref{rdSinest}).

\section{Causal structures in smooth product spacetimes} \label{c-s} 
\subsection{Smooth time-product spacetimes} 
Consider a $n$-dimensional spacetime $(V,\G)$ 
with base manifold the smooth product $V=I\times S$ where $I$ is an interval  
of $\R$, $S$ is  
a $n-1$ dimensional manifold and $t: V\rightarrow I$ the natural projection.  
If each slice $\{t\}\times S$ is spacelike as an embedded submanifold of $V$ 
and the vector field $\d/\d t$ is timelike  
(and will be assumed future-directed),  
then the metric $\G$ can be written globally as   
\be 
\G=\rho dt\otimes dt+{\bf \Omega}\otimes dt+ 
dt\otimes{\bf\Omega}-\h[t],\ t\in I,  
\label{timesproduct} 
\end{equation} 
where $\rho\in C^1(V)$ is positive, 
$\h[t]$ is a family of Riemannian metrics on $S$ and $\O$  
is a 1-form on $I\times S$. The resulting spacetime is called  
a {\em smooth time-product}. 
Notice that one can assume $I=\R$ without loss  
of generality, but the interval $I$ will be maintained some times for convenience.  
A first interesting case are {\em standard stationary} spacetimes,  
studied in subsection \ref{ex-stationary}.  
Another case, even more interesting, is when 
$\O=0$ which entails 
\be 
\G=\rho dt^2-\h[t],\ t\in I. 
\label{split} 
\end{equation}   
We shall employ the terminology {\em 1-timelike separable} spacetimes for  
these Lorentzian manifolds. Locally, any  spacetime can be written as in (\ref{split}),  
even with $\rho\equiv 1$  (the role of $\rho^{1/2}dt$ can be played by    
any integrable timelike 1-form);   
thus, expressions as (\ref{timesproduct}),  
(\ref{split}) are restrictive only from a global viewpoint.  
However, metrics such as (\ref{timesproduct}), and, especially,  
those which are 1-timelike separable,  
represent many physically interesting spacetimes and they arise in general  
settings. For instance, it has been recently shown that, for  
any globally hyperbolic spacetime, the metric tensor has the form (\ref{split}) 
\cite{BERNAL,BERNAL2}, and in this case 
$t$ can be set to a time function with Cauchy hypersurfaces $t=$const  (some extensions to stably causal  
spacetimes are also possible, see \cite{BERNAL2,SANCHEZ-BR}).  
\subsection{Arrival time functions} 
\label{time-arrival} 
It is not difficult to check in certain  
particular cases of (\ref{timesproduct}) whether there are curves with no  
particle horizons.  To that end, following \cite{BARI} we define the  
{\em future arrival} time  
function as the map $T^+:V\times S\rightarrow [0,\infty]$ given by   
$$ 
T^+((t_1,x_1),x_2)=\mbox{Inf}\{t-t_1:(t_1,x_1)\leq (t,x_2), t\in I\},\ \quad t_1\in I,\  
x_1,x_2\in S, 
$$ 
and dually for the past arrival  
time function $T^-$. Recall that if we define the {\em comoving trajectory}  
at $x_2$ as $R_{x_2}=\{(t,x_2):t\in I\}$, then, if $I=(a,b) \subseteq \R$:  
$$ 
\{t\in I:(t,x_2)\in R_{x_2}\cap I^+(t_1,x_1)\}  
= (t_1 + T^+((t_1,x_1),x_2), b). 
$$ 				 
Thus, the meaning of the arrival functions is the following: 
\bp 
In a smooth time-product spacetime 
we have  
$$ 
(t_1,x_1)\ll (t_2,x_2)\Longleftrightarrow 
T^+((t_1,x_1),x_2) < t_2-t_1 
\Longleftrightarrow 
T^-((t_2,x_2),x_1) < t_2-t_1. 
$$ 
\label{smooth-timeproduct} 
\ep 
The study of time arrival functions permits us to draw interesting 
conclusions about global causal properties of smooth time-product spacetimes;   
general properties and applications have been studied in \cite{SANCHEZ2, BARI}. In particular, 
 if the hypersurfaces  
$t=$const are Cauchy hypersurfaces then  
$T^{\pm}$ are continuous functions in their  
variables. Arrival time functions are related to the existence of  
particle horizons for comoving trajectories (for simplicity, we put $I=\R$). 
\bp 
Consider a smooth  
time-product spacetime $V=\R\times S$.  
Fixed $x_0$, the comoving trajectory  $R_{x_0}$   
has no past (resp. future) particle horizon if and only if  
$T^+(p,x_0)<\infty$, $\forall p\in V$ (resp. $T^-(p,x_0)<\infty$). 
\label{horizon-check} 
\ep 
 
\noindent 
\Pr According to proposition \ref{smooth-timeproduct}  
the condition on $T^+$ is clearly equivalent  
to $(t_1,x_1)\leq (t_0,x_0)$ for any point $p\equiv(t_1,x_1)$ and  
some $t_0\in\R$. 
\qed 
 
In particular, when $V_1, V_2$ are time-product spacetimes and $V_1 \prec_{\phi} V_2$ for a causal mapping which  
preserves the decomposition (\ref{timesproduct})  
(i.e., which maps each comoving trajectory  
$\{(t,x): t\in \R\}$ in $V_1$ into a comoving trajectory of $V_2$) then the finiteness of  
$T^+$ (resp. $T^-$) for $V_1$ implies the finiteness for $V_2$. 
Recall that sufficient conditions for the finiteness of $T^\pm$  
are easy to obtain \cite{BARI}  
(see also propositions \ref{psf}, \ref{psfGRW} below).

\subsection{Standard stationary spacetimes} 
\label{ex-stationary} 
 
A smooth time-product spacetime $I\times S$ as in  (\ref{timesproduct}) is called {\em standard stationary}  
if $I=\R$ and all the elements of $\G$ are independent of $t$, i.e.,  $\xiv=\d/\d t$ satisfies  
$\pounds_{\xiv}\rho=0$, $\pounds_{\xiv}\O=0$ and $\h[t] \equiv \h$. 
Locally, any stationary spacetime (i.e., a spacetime which admits a timelike Killing  
vector field $\xiv$) looks like a standard one.  
If the metric (\ref{timesproduct}) is both, standard stationary and {\em 1-timelike separable}  
then the spacetime is called {\em static standard} (see \cite{SANCHEZ-ST} for a survey). 
 
In such stationary $\R\times S$, any (non-constant)  
curve $\alpha$ contained in $S$ joining two fixed  
points $x_0, x_1$ yields a unique future-directed (resp. past-directed) lightlike curve $\g$ connecting a fixed $(t_0,x_0)$  
with some $(t_1,x_1)$, $t_1>t_0$  
 by means of the definition   
$\g:t\rightarrow (\tau(t), \alpha(t))$, $t\in [t_0,t_1]$ where  
$\tau(t)$ satisfies the differential equation  
$\rho^2\tau'^2+2\tau'{\bf\Omega}(\alpha')-\h(\alpha',\alpha')=0$ 
with $'\equiv d/d t$. 
Thus:  
\bp \label{psf} 
In a standard stationary spacetime, both arrival functions $T^+, T^-$ are always valued in $\R$. 
\ep 
Recall also that property {\em(\ref{point1})} of proposition  
\ref{set-properties} is satisfied  
in standard stationary spacetimes  
where, by definition, a foliation by achronal and spacelike hypersurfaces exists.  
These spacetimes do not satisfy property {\em(\ref{point2})}  
but they do satisfy property  
{\em(\ref{point4})}.  
 
\bp \label{pdifS} 
In a standard stationary spacetime $V= (\R\times S, \G)$, any smooth achronal  
hypersurface $\hat S$ which is closed (as a subset of $\R\times S$) is diffeomorphic to $S$. 
\ep 
 
\noindent 
\Pr 
As  
the vector field $\xiv$ is complete its flow defines a local diffeomorphism  
$\Psi: \hat{S} \rightarrow S$, which is injective by achronality.  
Its image is then an open subset $\Psi(\hat{S}) \subseteq S$. To  
check that it is closed and, thus, the equality holds, consider a  
sequence $\{x_m\}_{m=1}^{\infty}$ on $\Psi(\hat{S})$ which converges  
to a boundary point $x_0$. Take the sequence  
$\{\Psi^{-1}(x_m)\}_{m=1}^{\infty}\subset \hat{S}$ 
and define the quantities $T^{\pm}(\Psi^{-1}(x_1),x_{m})$, $m\in\N$. 
By proposition \ref{psf} all of them are finite and even more they are bounded 
by a constant independent of $m$. To see this last assertion,  
note that $T^{\pm}(\Psi^{-1}(x_1),x)\leq T^{\pm}(\Psi^{-1}(x_1),x_0)+C$ where $x$ lies  
in a neighbourhood of $x_0$ and $C$ is a constant. The main consequence of  
this is that all the values of $t$ for all the points of the sequence  
$\{\Psi^{-1}(x_m)\}_{m=1}^{\infty}$ 
are in a bounded interval which implies that, since $\hat{S}$ is closed,  
$\{\Psi^{-1}(x_m)\}_{m=1}^{\infty}$   
lies in a compact subset of $V$. Thus, it has a subsequence convergent  
to a point $\bar{x}_0\in \hat{S}$, and  
$x_0= \Psi (\bar{x}_0) \in \Psi(\hat{S})$, as required.  
\qed 
 
\bere \label{rdifS} 
{\em \ \\ 
\begin{enumerate} 
\item If $S$ is simply connected, the global condition ``achronal'' can be weakened to  
``locally achronal'' (i.e., with the induced metric not being  
Lorentzian at any point; obviously, this is fulfilled if $S$ is either spacelike or  
degenerate at any point), because this condition is enough to prove that $\Psi$ is  
a covering map (see theorem 4.4 of \cite{HARRIS-SHAPE} for a proof in a  
more general setting). Nevertheless, if $S$ is not simply connected the achronality  
cannot be weakened (just think in the Lorentzian cylinder, $\R \times S$, $S=S^1$,  
and take $\hat S$ as a spacelike helix). 
 
\item Remarkably, Harris and Low in \cite{HARRIS-SHAPE}  
proved a  more general result than proposition \ref{pdifS}:  
if a spacetime fulfills (i) $V$ admits a congruence ${\cal F}$ of  
inextensible timelike curves such that for any curve $\g\in{\cal F}$ we have 
that $I^\pm(\g)=V$, and (ii) there exist an achronal and 
properly embedded hypersurface $S$ in $V$, then  any other  
achronal hypersurface in $V$ is diffeomorphic to $S$  
(recall that ``properly embedded'' implies our assumption  
``closed as a subset''). 
A related result with the extra assumption of timelike or null  
geodesic completeness can be found in theorem 3 of \cite{GARFINKLE}. 
  
\end{enumerate} 
}\ere 
Notice that in de Sitter spacetime $\SSS_1^n$ the property stated  
in proposition \ref{pdifS} does not hold 
(example \ref{rdS}). 
Thus, as a consequence of  
proposition \ref{set-properties} one has  
the following result, applicable in particular when $V$ is Einstein static universe. 
 
\bc \label{cEidS} 
If $V$ is any standard stationary spacetime,  
$V\not\prec \SSS^n_1$. 
\ec

\subsection{General estimate for 1-timelike separable spacetimes} 
Next, we give a general estimate which ensures the existence of causal mappings  
between  1-timelike separable spacetimes. 
 We can assume that the base manifold is always the same and add the  
superscripts or subscripts  
$1$ and $2$ on the elements of the metric 
(\ref{split}) for each one of the  
two 1-timelike separable spacetimes. 
\bt 
Let $(V,\G_1)$, $(V,\G_2)$ be 
1-timelike separable spacetimes with respect  
to the same decomposition of $V$ written as $V=I\times S$. If $I$ is an  
unbounded interval then 
a sufficient set of conditions for  $\G_1\prec\G_2$  is the following:  
\begin{enumerate} 
\item  
$$ 
\begin{array}{c} 
k= \mbox{\large Inf}\\ 
\mbox{\footnotesize$t_1,t_2\in I,x\in S$} 
\end{array} \fr{\rho_2(t_2,x)}{\rho_1(t_1,x)}>0, 
$$ 
\item The norm $||\hat\h_2[t]||$  
of the endomorphism  
$\hat\h_2[t]$ \footnote{Recall that $\hat\h_2[t]$ can be regarded as a (self-adjoint) endomorphism on $T_xS$ and that this vector space is  
endowed with the Euclidean metric $\h_1[t]$ at $x$.  
Thus, by the {\em (pointwise) norm} we mean the standard Euclidean norm  
$||\hat\h_2[t]||^2 =$trace($\hat\h_2^2[t]$)  
(even though, alternatively, one can use, for example, the supremum norm).}  
with respect to $\h_1[t]$, defined in the tangent space $T_xS$ of any point  
$x\in S$ by the condition 
\be 
\h_2[t](\u,\v)=\h_1[t](\hat\h_{2}[t]\u,\v),\ \forall\u,\v\in  
T_xS 
\label{endomorphism} 
\end{equation} 
is bounded by a constant independent of $p\equiv(t,x)\in V$. 
\end{enumerate} 
\label{split-condition} 
\et 
\noindent 
\Pr  
This is proven by the explicit construction of a causal mapping  
$\Phi:(V,\G_1)\rightarrow (V,\G_2)$. We will perform the proof for $I=\R$  
but nothing essential changes if $I=]a,\infty[$ or $I=]-\infty,a[$ for some $a\in\R$. 
Define $\Phi$ by means of $\Phi:(t,x)\mapsto (l(t),x)$ where  
$l(t)$ is a strictly increasing $C^1$ function and $x\in S$. Then 
$$ 
\Phi^*\G_2=\rho_2(l(t),x)l'(t)^2dt^2-\h_2[l(t)]. 
$$ 
The endomorphism associated to $\Phi^*\G_2$ is in matrix form  
(naturally associated to (\ref{split})) 
$$ 
\widetilde{\Phi^*\G}_2=\left(\begin{array}{cc} 
\fr{\rho_2(l(t),x)}{\rho_1(t,x)}l'(t)^2 &\ \ \\ 
\ \ & \hat\h_2[t]\end{array}\right). 
$$ 
According to proposition \ref{DP-condition},  we deduce that 
$\Phi^*\G_2$ is a causal tensor if and only if 
\begin{equation} \label{eigen} 
\fr{\rho_2(l(t),x)}{\rho_1(t,x)}l'(t)^2\geq |\lambda_i(p)|,\  
i=1,\dots,n-1,  
\end{equation} 
where the $\lambda_i(p)$'s are the eigenvalues of  
$\hat\h_2[t]$.    
The condition on $||\hat\h_2[t]||$  
ensures that these eigenvalues will be functions of  
$p$ bounded by a constant  
$$ 
\begin{array}{ccc} 
N = &Sup&\!\!\!||\hat\h_2[t](x)||.\\ 
  &(t,x)\in V& \end{array} 
$$ 
On the other hand the inequality  
$$ 
\fr{\rho_2(l(t),x)}{\rho_1(t,x)}l'(t)^2>N, 
$$ 
which implies (\ref{eigen}), will hold whenever 
$$ 
l'(t)\geq\sqrt{\fr{N}{k}}, 
$$ 
in particular, by the choice $l(t)=(N/k)^{1/2}t$.\qed 
 
Interchanging the roles of $\G_1$ and $\G_2$, conditions for  
$\G_2\prec\G_1$ are obtained and, then: 
\bc 
Two 1-timelike separable spacetimes $(V,\G_1)$, $(V,\G_2)$, $V=I\times S$,  
written as in (\ref{split}) with unbounded $I$,   
are causally equivalent if, for some positive constants $N, N', k, k'>0$: 
$$k\leq \fr{\rho_1(t_2,x)}{\rho_2(t_1,x)} \leq  k',\   
\forall t_1, t_2 \in I,\ \forall x\in S,\quad   
N\leq ||\hat\h_2[t](x)||\leq N', \quad \forall (t,x)\in I\times S .$$ 
\label{split-corollary} 
\ec 
 
\bere \label{split-remark} 
{\em  
The results have been formulated with general functions $\rho$ to make them more easily applicable.  
Nevertheless, as the existence of causal mappings is a conformal invariant,   
the metric of (\ref{split}) can be rescaled by $1/\rho$, and  all the results re-formulated 
assuming that $\rho\equiv 1$. In this case we are only left with  
the second condition of theorem \ref{split-condition} and  
corollary \ref{split-corollary} and, in fact, {\em the so-obtained bounds are more general}.  
In a similar way, if $I$ were bounded then  
the change $t=f(\bar{t})$ with $\bar{t}$ ranging in an unbounded interval $J$ would  
bring the metrics into a form in which conditions of theorem \ref{split-condition} could 
be checked. 
} 
\ere

\subsection{GRW spacetimes} 
 
In previous subsection, we have obtained a general set of sufficient conditions for the  
causal equivalence of arbitrary timelike 1-separable spacetimes.  
Nevertheless, previous results (like those for stationary spacetimes or the example \ref{ejerectan})  
suggest the existence of many different causal structures, even in the globally hyperbolic case. 
To show this more explicitly, we focus now on  
a particular case of spacetimes. 
 
Generalized Robertson Walker spacetimes (GRW in short)  
are (1-timelike separable) warped products defined by:   
\be 
V= I \times_f S, \quad  \G=dt^2-f^2(t)\h, 
\label{grg} 
\end{equation} 
where $\h$ is a  Riemannian metric on the $(n-1)$-manifold $S$,   
and $f$ is a positive real function.  
Notice that the change 
\be 
T(t) =\int_{t_0}^t\fr{ds}{f(s)}, 
\label{integral} 
\end{equation} 
brings the above metric into the form  
\be 
\G =f^2(t(T))(dT^2-\h) 
\label{conf-flat} 
\ee 
where $T$ varies in a new interval, $I_T$. Thus, any GRW is conformal to a metric  
product; in particular, it is globally hyperbolic  
if and only if $(S,\h)$ is complete (see \cite{SANCHEZ-GRW} for further properties). 
The GRW spacetime will be called {\em spatially closed} 
if $S$ is compact (without boundary); recall that in this case the spacetime is globally hyperbolic.

Reasoning as for standard stationary spacetimes in proposition \ref{psf}, we have:  
\bp \label{psfGRW} 
In any GRW spacetime $I\times_fS$ with $I=\R$ and $f$ bounded, both  
arrival functions $T^+, T^-$ are always valued in $\R$. 
\ep 
(Clearly, the result still holds if $f$ only satisfies $\int_{t_0}^\infty 1/f= \infty,  
\int^{t_0}_{-\infty} 1/f= \infty$.) 
 
GRW spacetimes do not always satisfy  point {\em(\ref{point4})}  
of proposition \ref{set-properties}. In fact, de Sitter spacetime  
$\SSS^n_1$, which can be written  
as the spatially closed GRW spacetime $\R\times_{\cosh}\SSS^{n-1}$,  
is a counterexample (example \ref{rdS}).  
Nevertheless, the following result shows that the property is 
still satisfied in interesting cases. 
\bp \label{pdifSGRW} 
Consider a  spatially closed GRW  $I\times_fS$, and any smooth achronal  
hypersurface $\hat S$. Then, $\hat S$ is diffeomorphic to $S$ if one of  
the two following conditions hold: 
\begin{enumerate} 
\item $\hat S$ is compact. 
 
\item $I=\R$, $f$ is bounded and $\hat S$ is closed as a subset of $I\times S$. 
\end{enumerate} 
\ep 
 
\noindent 
\Pr Let $\Pi: V \rightarrow S$, $\Pi_R: V\rightarrow \R$ be the natural projections. 
\begin{enumerate} 
\item As  
the restriction of $\Pi$ to $S$ is a local diffeomorphism, necessarily the restriction  
$\Pi|_{\hat S}: \hat S \rightarrow S$ is a covering map. But 
the acausality of $S$ implies that this covering map has only one leaf, and hence  
it is a diffeomorphism. 
 
\item From the previous part, it is enough to prove that the hypotheses imply the compactness of $\hat S$. For any point $p\in\hat S$ the function defined by  
$T^+(p,\cdot): S \rightarrow [0,\infty]$ takes values in $\R^+$ (proposition \ref{psfGRW}) and, as it is continuous \cite[proposition 2.2]{SANCHEZ2},  its image is bounded in $\R$. 
 
The  acausality of $\hat S$,  
implies that the interval $\Pi_R(\hat S)$ is also bounded.  
To see this assume the contrary and   
 pick a point $p_1=(t_1,x_1)\in \hat S$ such that  
$|\Pi_R(p_1)-\Pi_R(p)|>T^\pm(p,x_1)$; this inequality means  
that there exists a timelike  
curve joining $p$ and $p_1$, which contradicts the achronality of $\hat S$.  
Therefore $\hat S$ lies in a compact subset of $V$. Since $\hat S$  
is closed, it is compact too, as required. 
\end{enumerate} 
\qed

\bere \label{rdifSGRW} 
{\em As in remark \ref{rdifS}, when $S$ is simply connected, ``achronality'' can be weakened to 
``local achronality''. 
}\ere 
 
Again, these results are applicable to de Sitter spacetime (and, in particular,  
for comparisons with Einstein static Universe, regarded as a GRW spacetime). 
 
\bc \label{cesu} 
If $V= \R\times_f \SSS^{n-1}$ is a GRW spacetime with $f$ bounded, then 
$V\not\prec \SSS^n_1$. 
\ec 
 
In order to obtain further conditions for the isocausality of spatially closed GRW,  
notice first that, as a consequence of corollary \ref{split-corollary}: 
\bl  
Two spatially closed GRW spacetimes $V_i=  I\times_{f_i}S$ with  
the same base manifold $I\times S$ and $I$ unbounded, are isocausal if  
$$         0< \mbox{ Inf}(f_i) \leq  \mbox{ Sup}(f_i) <\infty\ , \quad i=1,2. $$ 
\label{grw-corollary} 
\el 
 
\noindent 
\Pr Apply corollary \ref{split-corollary} taking into account that,  
for any point $(t,x)$,  
$$ 
\hat\h_2[t](x)=\fr{f_2(t)}{f_1(t)}\tilde{\alpha}_x, 
$$  
where $\tilde{\alpha}_x$ is the endomorphism associated to a (fixed) Euclidean scalar product  
 of the tangent $T_xS$ independent of $t$. So, the compactness of $S$  
yields the required inequality (\ref{eigen}) for the eigenvalues  
of $\hat\h_2$. \qed  
\bp \label{p5.5st} 
The causal structure of a spatially closed GRW spacetime $I\times_f S$ with $I$ unbounded and  
$0< \mbox{ Inf}(f) \leq  \mbox{ Sup}(f) <\infty$ is stable in the $C^0$ topology. 
\ep 
 
\noindent 
\Pr 
Let $\G$ be the warped metric, put $f_1= 2f, f_2=f/2$, and let $\G_i$ be the metric  
of the corresponding $I\times_{f_i} S$. The metrics with light cones strictly wider  
than $\G_1$ and strictly narrower  
than $\G_2$ constitute a $C^0$ neighbourhood of $\G$. Obviously, for any metric $\G'$ in such a  
neighbourhood $\G_1 \prec_{id} \G' \prec \G_2$ but, from lemma \ref{grw-corollary}, $\G_1 \sim \G_2$. 
\qed 
 
Of course proposition \ref{p5.5st} can be trivially extended to the case in which the intervals $I$ are not  
equal in both spacetimes, i.e., the base manifolds are $I_j\times S$, $j=1,2$,  
but both $I_j$ are unbounded with the same (upper, lower or both)  
infinite extremes. $S$ can also be replaced by  
two diffeomorphic compact manifolds $S_j$ but, essentially, no further generality is  
gained.  
Nevertheless, the restriction of the extremes being unbounded must hold.  
Let us see this necessity first in the simple case of product metrics.  
Notice that the completeness assumption for $\G_2$ in the following result is written  
only for simplicity, 
and holds automatically if $S$ is compact. 
 
\bl 
Consider the product spacetimes  
$$ 
V_1=(I_1\times S,\ \G_1=dt^2-\h_1),\  V_2=(I_2\times S,\ \G_2=dt^2-\h_2), 
$$ 
where $\h_1, \h_2$ are Riemannian metrics on $S$, $\h_2$ complete, 
and $I_1, I_2\subseteq\R$ are two open  
intervals. If $I_2$ is upper (resp. lower) bounded but $I_1$ is not 
then $\G_1\not\prec\G_2$. 
\label{lgrnogr} 
\el  
\noindent 
\Pr We only 
perform the proof for the case in which $I_2=]0,\infty[$, and  
$I_1=\R$ (the proof remains essentially equal for any other interval combinations). 
By proposition \ref{curve}, it is enough to show that there is an inextendible causal curve $\gamma$ in  
$V_1$ without particle horizon, whereas no such curve exist   
in $V_2$.  
In fact, from propositions \ref{horizon-check}, \ref{psfGRW}, the curve  
$\gamma(t)=(t,x_0), t\in \R$ in $V_1$ satisfies the required condition.  
To check the nonexistence of such a $\gamma$ for $V_2$, notice that, as $\g$  
would be causal, it can be reparametrized as $\gamma(t)=(t,x(t)), t>0$ with  
$\h_2(x'(t),x'(t)) \leq 1$. From the completeness of $\h_2$, there exists  
the limit lim$_{t\rightarrow 0}x(t) = x_0$. But $V_2$ can be regarded as an  
open subspace of $(\R\times S, dt^2-\h_2)$ and, then, $I^+(\gamma)= I^+(0,x_0) \neq I_2 \times S$. 
\qed 
 
As any GRW spacetime is conformally equivalent to a product one, combining the associate change of variable (\ref{integral}) with lemma \ref{lgrnogr} we have: 
\bp 
Consider two GRW spacetimes $I_i\times_{ f_i} S$ 
$$ 
V_1=I_1\times S,\ \G_1=dt^2-f_1^2(t)\h_1,\  V_2=I_2\times S,\  
\G_2=dt^2-f_2^2(t)\h_2, 
$$ 
where $I_1, I_2\subseteq \R$ are two open  
intervals $I_i=]a_i,b_i[$ and $\h_1$, $\h_2$ complete Riemannian  
metrics.  
Suppose also that $c_i\in I_i$ exists, such that one of the integrals 
$$ 
\int_{a_i}^{c_i} \frac{dt}{f_i(t)}, \quad \int_{c_i}^{b_i} \frac{dt}{f_i(t)}, 
$$ 
is infinite for $i=1$ and finite for $i=2$. Then  
 $\G_1\not\prec\G_2$. 
\label{grnogr} 
\ep

\bex 
\label{abierto}\em 
Consider the family of GRW spacetimes with $f(t)=t^a$, $a\in\R$,  
$t\in]0,\infty[$ and $S=\R^{n-1}$. Since 
\bnr 
\int_0^{c}t^{-a}dt<\infty,\ \mbox{if}\ a<1, 
\int_0^{c}t^{-a}dt=\infty,\ \mbox{if}\ a\geq 1,\\  
\int_c^{\infty}t^{-a}dt=\infty,\ \mbox{if}\ a\leq 1, 
\int_c^{\infty}t^{-a}dt<\infty,\ \mbox{if}\ a>1,  
\enr 
for any $c>0$ we deduce that spacetimes with $a<1, a=1, a>1$ are never isocausal. 
\eex  
 
Proposition \ref{grnogr} allows us to  
distinguish different causal structures in GRW spacetimes. When  
combined with lemma \ref{grw-corollary} and proposition \ref{p5.5st}, we can give a first  
classification of spatially closed GRW spacetimes.  
In order to give concrete physical examples, we will assume  
that the slices $t=$ constant are spheres, but the scheme works equally well  
for any type of compact slices. 
 
\bt \label{tclasifGRW} 
Consider any GRW spacetime $V= I\times_f S$ with $S$ diffeomorphic to a $(n-1)-$sphere. Then $V$ is isocausal to one and only one of the following four types of product spacetimes: 
\begin{enumerate} 
\item  
$\R \times  \SSS^{n-1}$, i.e.,  
Einstein static universe, with metric 
$$ 
\G=dt^2-d\O_{n-1}^2,\ t\in\R,    
$$ 
where $d\O_{n-1}^2$ represents the metric of the unit ($n-1$)-dimensional sphere.  
\item $]0,\infty[ \times \SSS^{n-1}$ with metric 
$$ 
\G=dt^2-\exp(2\alpha t)d\O^2_{n-1},\ t \in\R,\ \alpha<0. 
$$  
\item $]-\infty,0[ \times \SSS^{n-1}$. The metric is as  
in (ii) but now $\alpha>0$. 
 
\item $]0,L[ \times \SSS^{n-1}$, for some $L>0$.  
\end{enumerate} 
In the three first cases, the causal structure is  
$C^r$-stable in the set of all the metrics on $I\times S$.  
Moreover, causal structures belonging to the above cases 
can be sorted as follows 
$$ 
\mbox{\em coset}(\G_{iv})\preceq\left\{\begin{array}{c} 
              \mbox{\em coset}(\G_{ii})\\ 
               \mbox{\em coset}(\G_{iii})\end{array}\right\} 
\preceq\mbox{\em coset}(\G_{i}),   
$$ 
where the roman subscripts mean that the representing metric belongs  
to the corresponding point of the above description. 
\et 
 
\noindent 
\Pr Only the sorting of the causal structures remains to be proved.  
To that end we cast the representative metric of each causal structure  
in the form of (\ref{conf-flat}) obtaining 
\bnr 
\G_{i}=dt_1^2-d\O^2_{n-1},\ t_1\in]-\infty,\infty[\\ 
\fl\G_{ii}=\fr{1}{\alpha^2 t_2^2}(dt_2^2-d\O^2_{n-1}),\ t_2\in]0,\infty[,\  
\ \G_{iii}=\fr{1}{\alpha^2 t_3^2}(dt_3^2-d\O^2_{n-1}),\ t_3\in]-\infty,0[,\\  
\G_{iv}=dt^2_4-d\O^2_{n-2},\ t_4\in]0,L[. 
\enr 
>From these expressions it is not difficult to show that  
$\G_{iv}\prec\G_{iii}\prec\G_{i}$ and $\G_{iv}\prec\G_{ii}\prec\G_{i}$.  
Explicit causal mappings are (in all the cases only the time coordinate  
is involved) 
\bnr 
t_2=A\tan\left(\fr{\pi t_4}{2L}\right), t_3=A\tan\left(\fr{\pi(t_4-L)}{2L}\right), 
\ A\geq\fr{2L}{\pi},\\ 
t_1=Bt_2-\fr{A}{t_2},\ t_1=Bt_3-\fr{A}{t_3},\ A>0,\ B>1. 
\enr  
\qed

\bere \label{rdSinest} {\em Note that the first three  classes comprise each a single causal  
structure, whereas the fourth one contains more. To see it easily for $n=2$,  
consider example \ref{ejerectan}   
but instead of rectangles  
$R_L$ take the cylinders $C_L=]0,L[\times\SSS^1$.   
If $L< L'$ then $C_L \prec C_{L'}$,  
but the converse does not necessarily hold. In fact, for $L=\pi$ we have:  
$C_{\pi+\epsilon} \not\prec C_{\pi} \not\prec C_{\pi-\epsilon}$, for any  
$\epsilon \in ]0,\pi[$.  
This is so because, in $C_{\pi+\epsilon}$ there are timelike curves $\gamma$  
which satisfy  
$V=J^+(\gamma) \cup J^-(\gamma)$ (essentially property {\em(i)} of proposition  
\ref{future-common}), in $C_{\pi}$ no timelike curve satisfies this property,  
but  a lightlike curve $\gamma$ (in fact, any lightlike geodesic)  
satisfy $V=\overline{J^+(\gamma) \cup J^-(\gamma)}$, and in $C_{\pi-\epsilon}$  
no causal curve satisfies the property. 
This can be generalized to any dimension $n>2$. For example, if  
$C_L^n=]0,L[\times\SSS^{n-1}$ the relation $C^n_{\pi} \not\prec C^n_{\pi-\epsilon}$ follows because $C^n_{\pi}= J^-(J^+(\g))$ (resp. $C^n_{\pi-\epsilon}\neq J^-(J^+(\g))$, for any inextendible causal curve $\g$. 
}\ere 
Remarkably, de Sitter Universe  
$$ 
ds^2=dt^2-\cosh^2td\O^2_{n-1},\ t\in\R.    
$$ 
belong to this last class, with $L=\pi$. In fact, from (\ref{integral}), 
$$ 
L=\int_{-\infty}^{\infty}\frac{dt}{\cosh(t)}=\pi. 
$$ 
Notice that small modifications of $f(t)=\cosh(t)$ may change the value of the integral and,  
thus, the causal structure. Formally, recall that, as $\SSS^{n-1}$ is compact, any neighbourhood  
of the de Sitter metric for any $C^r$-Whitney topology must contain functions $f$ which satisfy, say, $f(t) \geq \cosh(t) $ (resp. $f \leq \cosh(t)$), $f(t_0)> \cosh(t_0)$  
(resp. $f(t_0)< \cosh(t_0)$)  
for some $t_0$, and $f=\cosh(t)$ outside a compact interval. Thus, the value of  
$L$ obtained for such a $f$ is smaller (resp. greater) than $\pi$ and, by remark  
\ref{rdSinest}, the corresponding spacetimes are not isocausal. Summing up: 
 
\bt \label{tdSinest} 
For any neighbourhood ${\cal U}$ in a $C^r$-Whitney topology, $r=0,1,\dots , \infty$ of de Sitter spacetime, there is a spacetime $V \in {\cal U}$ such that 
$$V \not\sim \SSS^n_1 .$$ 
Thus, the causal structure of de Sitter spacetime $\SSS^n_1$ is {\em unstable.} 
\et

\section{Mp-waves} \label{smp} 
 
\subsection{General results} 
Mp-waves are $n$-dimensional  
Lorentzian manifolds whose topology is that of a product $V=\R^2\times M$, 
where $M$ is a connected manifold endowed with a Riemannian metric $\h$. 
 If we set a global coordinate chart on the Lorentzian manifold defined by $z=\{u,v,x\}$  
with $\{u,v\}$ canonical coordinates for $\R^2$ and $x=\{x^1,\dots,x^{n-2}\}$, 
the coordinates of $M$ the Lorentzian metric $\G$ is then 
\be 
\G=2dudv-\h[u]+H(x,u)du^2,  
\label{mp-wave} 
\end{equation} 
where $\h[u]$ is the Riemannian metric alluded to  
above (note that it depends explicitly on $u$) 
Here the scalar function $H(x,u)$ is in principle $C^0$ although one may  
need to add higher differentiability conditions on it according to the problem under  
study. The nomenclature used here for these spaces is not standard but we feel  
that it is less misleading than the traditional one ``plane fronted waves with parallel 
rays'' or in short $pp$-waves. This is so because the spaces defined by (\ref{mp-wave})  
admit a covariantly constant lightlike vector field  
(this is the vector $\d/\d v$ in our coordinates) so they certainly contain  
parallel rays but in general the wave fronts ($u=const$) are not planes (see  
\cite{REVIEW} for a further discussion).  
 
Particular cases of Mp-waves have received wide attention recently particularly  
by the string theory community. For us though, studies dealing  
with the global 
causal properties of these Lorentzian manifolds will be more relevant and in fact  
since the classical work of Penrose \cite{PENROSE-WAVE} great progress has been made.   
The most researched Mp-waves are those in which the Riemannian metric does not  
depend on $u$, and we shall drop the letter $u$ from $\h$ in this case (recall that the name {\em PFW} has also been used in this case, \cite{CANDELA}). For such Mp-waves, 
a very general classification of their causal 
properties was accomplished  in terms of the asymptotic behaviour of 
 $H(x,u)$ in the variable $x$,  \cite{SANCHEZ}. Other relevant aspects which 
have been studied for these Lorentzian manifolds are the construction of the  
{\em causal boundary} for certain particular cases of $H(x,u)$  
\cite{MAROLF-ROSS,MAROLF-ROSS2,HUBENY}, the presence of event horizons \cite{HR,TRAP},  
\cite[Sect. 3.2]{FS} or  
their geodesic connectivity \cite{CANDELA}.       
 
In this subsection we will show how our methods provide a simple way to group  Mp-waves 
in sets with the same causal structure. To that end let us agree to call $H_1(x,u)$, 
$H_2(x,u)$, $\h_1[u]$, $\h_2[u]$ the scalar functions and Riemannian metrics of two different 
Mp-waves with the same base manifold. Next result establishes very simple relations between 
these objects in order that the Mp-waves they represent be causally related.  
 
\bt 
The Mp-waves $\G_1$ and $\G_2$ represented by $\h_1[u]$, $H_1(x,u)$, $\h_2[u]$, $H_2(x,u)$  
are causally related ($(V_1,\G_1)\prec(V_2,\G_2)$)  
if the following conditions are met 
\begin{itemize} 
\item we can find strictly positive constants $k_1$, $k_2$  
satisfying the inequality 
\be 
k_2H_2(x,u_2)-k_1H_1(x,u_1)\geq 0,\ \forall u_1, u_2\in\R,\ \forall x\in M. 
\label{condition1} 
\end{equation} 
\item The endomorphism 
$\hat \h_2[u]$  
defined in the tangent space of each point $z$ 
by the condition  
\be 
\h_1[u](\v_1,\hat\h_2[u]\v_2)=\h_2[u](\v_1,\v_2),\ \forall\v_1,\v_2\in T_z(V), 
\label{condition2} 
\end{equation} 
has its norm $||\hat\h_2[u]||$, when regarded as function of $z$, 
bounded from above by a constant. 
\end{itemize} 
\label{mp-result} 
\et

\noindent 
\Pr To show this result we construct an explicit causal mapping from $(V_1,\G_1)$ onto  
$(V_2,\G_2)$. In the coordinates of (\ref{mp-wave}) define the diffeomorphism  
$\Phi:(u,v,x)\mapsto (f(u),g(v),x)$ for certain differentiable and monotone functions  
$f(u)$, $g(v)$. The pull-back $\Phi^*\G_2$ is then 
$$ 
\Phi^*\G_2=2f'(u)g'(v)dudv+H_2(x,f(u))f'(u)^2du^2-\h_2[f(u)]. 
$$ 
>From this we can easily calculate the endomorphism associated to $\Phi^*\G_2$ 
(see subsection \ref{causal-tensor}) which in the natural basis used in  
(\ref{mp-wave}) takes the form  
$$ 
\left(\begin{array}{ccc} 
f'(u)g'(v) &\ 0 \ &\ 0 \\ 
f'(u)^2H_2(x,f(u))-H_1(x,u)f'(u)g'(v) & f'(u)g'(v) &\ 0\\ 
\ 0 \ &\ 0 \ & \hat\h_2[u] \end{array}\right).  
$$ 
This endomorphism has the algebraic type explained in 
the second point of proposition \ref{DP-condition} 
and so it is causal-preserving if the conditions 
$$ 
f'(u)H_2(x,f(u))-g'(v)H_1(x,u)\geq 0,\ f'(u)g'(v)\geq |\lambda_j(z)|, 
$$ 
hold, where $\{\lambda_j(z)\}, j=1, \dots, n-2$ are the eigenvalues of the endomorphism $\hat\h_2[u]$; notice that they are bounded by a constant, namely $a^2, a>0$, as functions of $z$.  
Under our hypotheses these inequalities are clearly 
fulfilled if we take $f(u)=a k_2u$, $g(v)=a k_1v$ proving that such  
$\Phi$ is a causal mapping. 
\qed  
 
\bere 
\noindent {\em  
This proposition also supplies sufficient  conditions for the isocausality of Mp-waves, by interchanging the roles of the labels $1$ and $2$. Note that, then,  equation (\ref{condition2}) would  
define a new endomorphism  
$\hat\h_1[u] =(\hat\h_2[u])^{-1}$.   
} 
\ere 
 
\subsection{Application to plane waves} 
Theorem \ref{mp-result} can be applied in a number of interesting particular  
cases as we detail next. If the wave fronts are planes ($\h[u]$ is flat for any  
fixed $u$) 
then the resulting $pp$-wave  
can be further classified according to the scalar function as follows: 
\begin{enumerate} 
\item Plane waves: these are plane fronted waves with further isometries aside from 
the vector field $\d/\d v$ the wave fronts being hypersurfaces of transitivity.  
In the coordinates of (\ref{mp-wave}) the metric tensor takes the form 
$$ 
\G=2dudv+ \sum_{i,j=1}^{n-2} A_{ij}(u)x^ix^j \, du^2-h_{ij}dx^idx^j, 
$$ 
where $h_{ij}$ are constants representing a symmetric positive definite bilinear form. 
The matrix $A_{ij}(u)$ is called the {\em frequency matrix}. 
 
\item Locally symmetric plane waves: this is a 
particular case of the above in which the frequency matrix does not depend on  
$u$. The curvature tensor of these metrics is covariantly constant and this motivates  
the terminology, although names such as {\em homogeneous plane waves} can be also  
found in the literature (we have avoided this last terminology because it is  
sometimes used for more general plane waves \cite{BLAU}).  
\end{enumerate} 
 
Let us consider first the latter case. According to above considerations, for a locally symmetric plane wave 
 canonical coordinates in which (\ref{mp-wave}) takes the form 
\be 
\G=2dudv+ \sum_{i=1}^{n-2}\epsilon_i \, (x^i)^2 \, du^2-h_{ij}dx^idx^j 
\label{h-w} 
\end{equation}  
 can always be found. 
Here $\epsilon_i=\pm 1$  
\footnote{Eventually, one must also admit $\epsilon_i=0$  
if the frequency matrix is degenerate, but we will not deal  
with this case.} $\forall i=1,\dots,n-2$. Alternatively we can bring the  
Riemannian part into its diagonal form by means of a linear transformation obtaining 
\be 
\G=2dudv+Q(x,x)du^2-\sum_{i=1}^{n-2}(dx^i)^2, 
\label{canonical} 
\end{equation} 
where $Q(\cdot,\cdot)$ is a symmetric bilinear form with signature 
given by the set $\{\epsilon_i\}$. 
We will use (\ref{h-w}) or (\ref{canonical}) in accordance with the problem under  
study. Theorem \ref{mp-result}, applied  to pairs of  
locally symmetric plane waves, yields: 
 
\bp 
Two locally symmetric plane waves $(V_i,g_i), i=1,2$ with scalar functions $H_i= Q_i$ of the same signature 
as quadratic forms, are always causally equivalent. 
\label{quadric-equivalent} 
\ep 
 
\noindent 
\Pr 
It is clear that in this case the endomorphism $\hat\h_2[u]$  
is just a constant linear mapping from $\R^{n-2}$ to $\R^{n-2}$,  
independent of $z$.  
Hence the second condition of  
theorem \ref{mp-result} is automatically satisfied (either if  
the representation (\ref{h-w}) or (\ref{canonical}) is chosen). 
For the first one, since $Q_1$, $Q_2$ have the same signature, the coordinates of (\ref{h-w}) can  
be chosen in such a way that $Q_1$, $Q_2$ yield the same 
quadratic form, $Q$ on $\R^{n-2}$ (the Riemannian  
parts will be different for each metric). Thus, condition  
(\ref{condition1}) will be satisfied by just putting $k_1=k_2=1$. 
\qed 
 
Now, let us see that the conditions of theorem \ref{mp-result}  also hold for other types of  
plane waves. As before, the endomorphism $\hat\h_2[u]$ is just a linear mapping and hence $||\hat\h_2[u]||$ 
is constant so only condition (\ref{condition1}) must be studied. Denoting by  
$A^1_{ij}(u)$, $A^2_{ij}(u)$ the frequency matrices of each plane wave it is clear that  
this condition (for $\G_1\prec\G_2$) entails 
\be 
k_2A^2_{ij}(u)x^ix^j-k_1A^1_{ij}(u)x^ix^j\geq 0,\ (x^1,\dots,x^{n-2})\in\R^{n-2},\ u\in\R.\  
\label{p-inequality} 
\end{equation} 
This inequality will hold if and only if the quadratic form $kA^2(u)-A^1(u)$ is  
semidefinite positive for some $k=k_2/k_1 >0$.  
If both $A^1(u)$, $A^2(u)$ are positive definite, we deduce that  
at each $u$, $k$ would satisfy $k > \lambda_{\hbox{max}}^1(u)/\lambda_{\hbox{min}}^2(u)$, where  
$\lambda_{\hbox{max}}^i(u)$ (resp. $\lambda_{\hbox{min}}^i(u)$)  
denotes the maximum (resp., minimum) of the eigenvalues of $A^i(u)$.  
In order to include the case in which $A^1(u)$, $A^2(u)$ are negative definite,  
we must regard  
$\lambda_{\hbox{max}}^i(u), \lambda_{\hbox{min}}^i(u)$ as the maximum or  
minimum of the absolute values of the corresponding eigenvalues: 
 
\bp 
Two planes waves,  with frequency matrices $A^1(u)$, $A^2(u)$,  
either both positive definite  
or both negative definite  
are isocausal if: 
 
$$ \mbox{Sup}_{u\in \R} \{\lambda_{\mbox{max}}^1(u)/\lambda_{\mbox{min}}^2(u)\} < \infty \quad 
\quad \mbox{Sup}_{u\in \R} \{\lambda_{\mbox{max}}^2(u)/\lambda_{\mbox{min}}^1(u)\} < \infty .$$ 
\label{pol} 
\ep

\subsection{Causal boundaries of plane waves} 
\label{ppwaveb} 
 
Theorem \ref{quadric-equivalent} enables us to construct  
explicitly in certain plane waves the causal embedding boundary put forward  
in definition  
\ref{cboundary}. To see how this is achieved let us 
consider the case of locally symmetric plane waves.  
The Weyl tensor of these spacetimes  
($n\geq 4$) can be explicitly calculated and, in the  
coordinates of (\ref{canonical}), its only nonvanishing components are 
$$ 
C_{ux^iux^j}=-Q_{ij}+\fr{1}{n-2}\delta_{ij}\sum_{k=1}^{n-2}Q_{kk}. 
$$ 
This implies that the plane wave (\ref{canonical})  
is conformally flat if and only if $Q_{ij}=\lambda\delta_{ij}$,  
with $\lambda=\sum_{k=1}^{n-2}Q_{kk}/(n-2)$.   
These are particular  
cases of proposition \ref{quadric-equivalent}  
which in fact include a bigger  
class of locally symmetric plane waves non-conformally flat and  
whose scalar function has definite sign. Now the  
conformal boundary of conformally flat locally symmetric plane waves  
can be constructed explicitly and hence the conformal embeddings needed 
will turn into causal extensions for any of the causally equivalent  
cases studied in proposition \ref{quadric-equivalent}. 
We summarize next  the known results on  
conformal boundaries  for  
locally symmetric plane waves and for the sake of completeness    
we also give account of other notions of causal boundary 
valid for non-conformally flat ones.  
 
\begin{enumerate} 
\item {\em Conformally flat case.}  
We must distinguish between $\lambda>0$ or $\lambda<0$ 
\begin{itemize} 
\item[A] $\lambda>0$.  An explicit conformal embedding 
in dimension $n=10$ into  
Einstein static universe is claimed in \cite{BERENSTEIN}. 
The conformal boundary is a null one-dimensional line.  
\item[B] $\lambda<0$. The conformal completion for this  
case was known since long ago   
and it turns out that the Lorentzian manifold is conformally  
related to a region of $\LLL^n$  
bounded by two lightlike planes. 
\end{itemize} 
 
\item {\em Non-conformally flat case.} $Q$ may have any signature.  
A causal boundary when the matrix 
$Q_{ij}$ has at least a positive definite eigenvalue 
has been constructed in \cite{MAROLF-ROSS}.  
They showed that this boundary can be again regarded as  
a one-dimensional line.  
As far as we know, there are no known results for other cases. 
\end{enumerate} 
 
\smallskip 
 
\smallskip 
 
\noindent  Now, recall that proposition \ref{quadric-equivalent}  
tells us that any  (conformally flat or not) locally symmetric  
plane wave $(V,\G)$ with $Q$ either positive or negative definite  
is isocausal to one of the cases (1A), (1B). Thus,  the conformal  
boundary obtained in each one of these cases  
is a causal embedding boundary in the sense of definition \ref{cboundary} for $V$ 
with $i=i_1\circ i_2$ where $i_2$ is a causal mapping from $V$ to the  
manifold of the corresponding case (1A) or (1B) and $i_1$ the conformal  
embedding.  
In particular, this holds for the case with $Q$ negative definite and  
non-conformally flat. We remark that, applying the results of  
\cite{SANCHEZ} one can  
prove that such plane waves are always globally hyperbolic;  
this matches the result that the causal embedding boundary constructed  
here is formed by two  
lightlike planes limiting a sandwich region of $\LLL^n$. 
 
These considerations can be extended to any plane wave isocausal to  
a locally symmetric plane wave with the above properties. For instance if the 
frequency matrix is negative definite then proposition \ref{pol} establishes 
that this plane wave is isocausal to a locally symmetric plane 
wave of the type (1B) if the eigenvalues of the frequency matrix  
fulfill the condition 
$$ 
0<\lambda_i(u)<\infty,\ \forall i, 
$$  
as is very easy to check. Therefore our causal embedding boundary for these plane  
waves is formed by two lightlike planes in the same fashion as before. 
 
\section*{Acknowledgements} 
A.G.P. wishes to thank the Departmento de Geometr\'{\i}a y Topolog\'{\i}a 
of Universidad de Granada  (Spain) 
for funding a short term visit during which this work was  
developed and the warm hospitality displayed.  
A.G.P. also acknowledges the financial support of  
the research grants BFM2000-0018 and FIS2004-01626 of the Spanish CICyT and  
no. 9/UPV 00172.310-14456/2002 of the Universidad del Pa\'{\i}s Vasco.  
M.S. is partially supported by MCyT-FEDER grant 
$\textrm{n}^{\circ}$ MTM 2004-04934-C04-01.

Both authors thank Jos\'e M. 
M. Senovilla for a careful reading of the manuscript and  
his many improvements and suggestions. The comments of two anonymous  
referees are also gratefully  acknowledged.

\end{document}